\documentclass[10pt,amsmath,eqsecnum,showpacs,showkeys]{revtex4}
\newcommand{\eq}[1]{Eq.~{(\ref{#1})}}
\newcommand{\fig}[1]{Fig.~{\ref{#1}}}
\newcommand{\bea}{\begin{eqnarray}}
\newcommand{\beann}{\begin{eqnarray*}}
\newcommand{\eea}{\end{eqnarray}}
\newcommand{\eeann}{\end{eqnarray*}}

\usepackage{graphicx}
\usepackage{latexsym}
\usepackage{amssymb}
\usepackage{amsbsy}
\DeclareMathAlphabet{\mathsfsl}{OT1}{cmss}{m}{sl}
\newcommand{\bs}{\boldsymbol}
\renewcommand{\vec}[1]{\bs{#1}}
\newcommand{\hvec}[1]{\hat{\vec{#1}}}
\newcommand{\mv}[1]{\mathbf{#1}}

\newcommand{\pv}[1]{\underline{\vec{#1}}}
\renewcommand{\tensor}[1]{\mathsfsl{#1}}
\newcommand{\op}[1]{\underline{\tensor{#1}}}
\newcommand{\inv}[1]{\op{#1}^{-1}}
\newcommand{\adj}[1]{\overline{\tensor{#1}}}
\newcommand{\adjinv}[1]{\adj{#1}{}^{-1}}
\newcommand{\Cl}{\mathcal{C}\ell}
\newcommand{\pd}{\vec{\partial}}
\newcommand{\dd}{\mathrm{d}}
\newcommand{\DD}{\tensor{d}}
\newcommand{\grade}[1]{\left\langle #1 \right\rangle}
\newcommand{\grad}{\stackrel{\rightharpoonup}{\vec{\nabla}}}

\begin{document}
\title{Geometric Algebra Techniques for General Relativity}
\author{Matthew R. Francis} \email{mfrancis@physics.rutgers.edu}
\author{Arthur Kosowsky} \email{kosowsky@physics.rutgers.edu} \affiliation{Dept. of Physics and Astronomy, Rutgers University\\136 Frelinghuysen Road, Piscataway, NJ 08854}
\date{\today}

\begin{abstract}
Geometric (Clifford) algebra provides an efficient mathematical language for describing physical problems.  We formulate general relativity in this language.  The resulting formalism combines the efficiency of differential forms with the straightforwardness of coordinate methods.  We focus our attention on orthonormal frames and the associated connection bivector, using them to find the Schwarzschild and Kerr solutions, along with a detailed exposition of the Petrov types for the Weyl tensor.
\end{abstract}
\pacs{02.40.-k; 04.20.Cv}
\keywords{General relativity; Clifford algebras; solution techniques}
\maketitle

\section{Introduction}

Geometric (or Clifford) algebra provides a simple and natural language
for describing geometric concepts, a point which has been argued
persuasively by Hestenes \cite{STA} and Lounesto \cite{lounesto} among
many others. Geometric algebra (GA) unifies many other mathematical formalisms
describing specific aspects of geometry, including complex variables,
matrix algebra, projective geometry, and differential
geometry. Gravitation, which is usually viewed as a geometric theory,
is a natural candidate for translation into the language of geometric
algebra.  This has been done for some aspects of gravitational theory;
notably, Hestenes and Sobczyk have shown how geometric algebra greatly
simplifies certain calculations involving the curvature tensor and provides techniques for classifying the Weyl tensor \cite{CA2GC,STAC}.
Lasenby, Doran, and Gull \cite{GTG} have also discussed gravitation using
geometric algebra via a reformulation in terms of a gauge principle.

In this paper, we formulate standard general relativity in terms of
geometric algebra.  A comprehensive overview like the one presented
here has not previously appeared in the literature, although
unpublished works of Hestenes and of Doran take significant steps in
this direction.  We obtain several useful results not available in
standard treatments, including a general solution for the connection
in terms of coordinate frame vectors (Section~\ref{GR_in_GA}) and simplified
forms for the Petrov classification of the Weyl tensor (Section~\ref{petrov_class}). Various mathematical
objects, such as the connection and the curvature tensor, gain
explicit geometrical interpretations through their mathematical forms. 
Also, geometric algebra provides notable compactness and clarity
for many derivations, such as the generalized
Schwarzschild solution in Section~\ref{schwarz_geom} and the Kerr solution in Section~\ref{kerrsol}.  An explicit translation between geometric algebra and
differential forms is provided in Section~\ref{diffgeom}, which provides
orientation to those unfamiliar with geometric algebra but conversant
in conventional differential geometry.

We assume knowledge of Clifford algebra techniques and notation, and their geometric interpretation. Appendix~\ref{intro_CA} reviews basic elements and lists some pedagogical references, while Appendix~\ref{intro_GC} defines basic derivative operators and tensors. Definitions and results from these Appendices will be used without comment.

\section{Covariant Derivatives and Curvature}
\label{GR_in_GA}

\subsection{Orthonormal Frames and The Bivector Connection}

Since we are working within the framework of standard general relativity, we will assume the existence of the spacetime manifold, and the presence of a metric on it.  This means we can define the covariant derivative $\nabla$ in the standard way (see Section 3.1 of \cite{wald}), so that it (1) is linear, (2) obeys the Leibniz rule, (3) commutes with contraction, and (4) reduces to $\vec{\partial}$ in its action on scalar functions (see Appendix~\ref{intro_GC} for details); we also impose the torsion-free condition, which we will discuss in more detail later.  We specify the operator uniquely by letting it be metric-compatible.

The scalar-valued derivative operator $\vec{t}\cdot\nabla$ can be decomposed into the form
\bea
\vec{a} \cdot \nabla \mv{A} \equiv \vec{a} \cdot \hat{\nabla} \mv{A}
+ \tensor{\Gamma}(\vec{a},\mv{A}),
\label{cov_decomp} 
\eea
where $\tensor{\Gamma}$ is called a \emph{connection}, which has the same grade (or grades) as $\mv{A}$.  The symbol $\hat{\nabla}$ is another derivative operator that we can choose:  for example, if we want $\tensor{\Gamma}$ to be the Christoffel connection, we let $\hat{\nabla} \equiv \vec{\partial}$.  We will instead adapt the covariant derivative to a particular frame of vectors, whose ``metric'' is fixed from point to point in the manifold.

A frame of vectors $\{\vec{e}_\mu\}$ is one for which the pseudoscalar 
(volume element) does not vanish,
\bea
e = \vec{e}_0 \wedge \vec{e}_1 \wedge \vec{e}_2 \wedge \vec{e}_3 \neq 0,
\eea
for a region of spacetime; the existence of a pseudoscalar in some region is equivalent to the existence of a metric on the same region of a manifold \cite{CA2GC}.  (See also \eq{duality} in Appendix~\ref{intro_CA}.)  A given vector in the reciprocal frame is found via
\bea
\vec{e}^\mu = (-1)^\mu \bigwedge\limits_{\nu \neq \mu} \vec{e}_\nu e^{-1},
\eea
which satisfies the orthonormality condition
\bea
\vec{e}^\mu \cdot \vec{e}_\nu = \delta^\mu_\nu .
\eea
Our starting point in developing a geometric algebra description of
spacetime is a particular class of frame vectors, the orthonormal frame
$\{\hvec{e}^\mu\}$ satisfying
the condition
\bea 
\hvec{e}^\mu \cdot \hvec{e}^\nu = \eta^{\mu\nu},
\label{ortho_frame} 
\eea 
where $[\eta]=\text{diag}(1,-1,-1,-1)$ is the metric for Minkowski spacetime.  (In fact, we do not need the frame to be orthonormal---all we need is for the metric in this frame to be constant.)  We attach this orthonormal frame to each point of the spacetime manifold, but we will leave the discussion of metrics and coordinate frames to Section~\ref{frames}.

If we apply the covariant derivative to \eq{ortho_frame} we find the special property
\bea
	\vec{a} \cdot \nabla (\hvec{e}_\mu \cdot \hvec{e}_\nu ) = \vec{a} \cdot \pd (\hvec{e}_\mu \cdot \hvec{e}_\nu ) = 0 ;
\label{constant_frame}
\eea
applying the Leibnitz rule over the inner product yields
\bea
\vec{a} \cdot \nabla (\hvec{e}_\mu \cdot \hvec{e}_\nu ) = \left( \vec{a}\cdot \nabla \hvec{e}_\mu \right) \cdot \hvec{e}_\nu + \hvec{e}_\mu \cdot \left( \vec{a}\cdot \nabla \hvec{e}_\nu \right) \equiv \Gamma_{\vec{a}}(\hvec{e}_\mu , \hvec{e}_\nu ) + \Gamma_{\vec{a}}(\hvec{e}_\nu , \hvec{e}_\mu ) = 0
\eea
where $\Gamma$ is a representation of the connection components with respect to the basis $\{\hvec{e}_\mu\}$.  Since the function $\Gamma_{\vec{a}}(\hvec{e}_\mu , \hvec{e}_\nu ) = - \Gamma_{\vec{a}}(\hvec{e}_\nu , \hvec{e}_\mu )$ is antisymmetric and linear in its arguments, we can write it in terms of a bivector as follows (as shown in Chapter 3 of \cite{CA2GC}):
\bea
	\Gamma_{\vec{a}}(\hvec{e}_\mu , \hvec{e}_\nu ) \equiv \mv{\Omega}(\vec{a}) \cdot ( \hvec{e}_\mu \wedge \hvec{e}_\nu ) = \left( \mv{\Omega}(\vec{a}) \cdot \hvec{e}_\mu \right) \cdot \hvec{e}_\nu ,
\eea
where $\mv{\Omega}(\vec{a})$ is the \emph{bivector connection}.  This means the covariant derivative of a frame vector can be written as
\bea
\vec{a}\cdot \nabla \hvec{e}_\mu \equiv \mv{\Omega}(\vec{a}) \cdot \hvec{e}_\mu ,
\eea
while for an arbitrary multivector we have the decomposition
\bea
	\vec{a} \cdot \nabla \mv{A} = \vec{a} \cdot \hat{\nabla} \mv{A} + \mv{\Omega}(\vec{a}) \times \mv{A},
\label{cov_deriv}
\eea
where we have imposed the condition $\vec{a} \cdot \hat{\nabla} \hvec{e}_\mu = 0$ for frame vectors and $\vec{a} \cdot \hat{\nabla} \phi = \vec{a} \cdot \pd \phi$ for scalars; when the frame vectors $\hvec{e}_\mu$ are constant, $\hat{\nabla} = \pd$.  (Note that $\times$ is the commutator product, not the vector cross-product:  $\mv{A}\times\mv{B} = (\mv{AB}-\mv{BA})/2$.)  Due to the ``fixed'' nature of the orthonormal frame, the connection bivector acts as a ``rotation'' of objects with respect to a ``background'' defined by the $\{\hvec{e}_\mu\}$; this is the foundation of the flat-spacetime gauge theory of gravity expounded by Lasenby and his colleagues \cite{GTG,SCGT}.  The relation to gauge theory is given in a note following this section.

With the covariant derivative in hand, we define parallel transport
in the usual way; the magnitude of the vector being transported is constant, so
\bea
\vec{t} \cdot \nabla \vec{b}^2 = 0 = 2 \vec{b} \left( \vec{t} \cdot \nabla \vec{b} \right) \quad \to \quad \vec{t} \cdot \nabla \vec{b} = 0.
\eea
A \emph{geodesic} is defined when the tangent vector is parallel transported along itself:
\bea
\vec{t} \cdot \nabla \vec{t} = 0 .
\eea

We can write three derivative operators based on \eq{cov_deriv}:
\begin{subequations}
\bea
\nabla \cdot \mv{A} &=& \vec{\partial_a} \cdot \left( \vec{a} \cdot \nabla \mv{A} \right) = \hat{\nabla} \cdot \mv{A} + \vec{\partial_a} \cdot \left( \mv{\Omega}(\vec{a}) \times \mv{A} \right) \qquad \text{covariant divergence} \\
\nabla \wedge \mv{A} &=& \vec{\partial_a} \wedge \left( \vec{a} \cdot \nabla \mv{A} \right) = \hat{\nabla} \wedge \mv{A} + \vec{\partial_a} \wedge \left( \mv{\Omega}(\vec{a}) \times \mv{A} \right) \qquad \text{covariant curl} \\
\nabla \mv{A} &=& \nabla \cdot \mv{A} + \nabla \wedge \mv{A} = \vec{\partial_a} \left( \vec{a} \cdot \nabla \mv{A} \right) \qquad \text{covariant gradient},
\eea
\end{subequations}
where $\vec{\partial}_a$ is defined in Appendix~\ref{intro_GC}.  As we will see in Section~\ref{diffgeom}, the covariant curl is equivalent to the exterior derivative of differential geometry (when torsion is zero).

\subsection{Geometric Interpretation of the Connection}

Comparison of two infinitesimally close points in curved spacetime is not as simple as its analogue in a flat space, but if we assume that the orthonormal frame may be carried smoothly from one point to another in the manifold, many of the results of flat spacetime carry over.  The same multivector field at different points $\mv{A}(\tau)$ and $\mv{A}(\tau')$ along a curve $\tensor{f}$ parameterized by some parameter $\tau$ will not in general have the same magnitude \emph{or} direction.  The covariant derivative is a measure of this difference, which can be viewed geometrically as the difference between the multivector at a
given point on a curve and its value at an infinitesimally displaced
point along the curve which has been parallel transported (that is, transported without changing its magnitude or direction) back to the
original point.  Without loss of generality we evaluate the derivative at $\tau = 0$:
\bea
\vec{t} \cdot \nabla \mv{A} = \lim_{\tau \to 0} \frac{1}{\tau} \left( \inv{\Lambda}_{(-\tau)} (\mv{A}_\tau) - \mv{A}_{\tau = 0} \right) = \left.\frac{\dd}{\dd\tau} \inv{\Lambda}_{(-\tau)} (\mv{A}_\tau)\right|_{\tau = 0}
\label{co_param}
\eea
where
\beann 
\vec{t}[\tensor{f}] \equiv \frac{\dd \tensor{f}(\tau)}{\dd \tau}
\eeann
is a vector tangent to the curve and $\inv{\Lambda}_{(-\tau)} (\mv{A}_\tau)$ is the parallel-transported multivector evaluated at $\tau=0$.  We are only able to define $\inv{\Lambda}$ in either a flat space, or if we are able to carry the orthonormal frame smoothly along the curve $\tensor{f}$; see \cite{choquet}, Chapter Vbis for more information.

We can simplify \eq{co_param} by letting $\inv{\Lambda}_{\tau = 0} = \op{1}$ (the identity operator) and $\mv{A}(\tau = 0) = \mv{A}$:
\bea
\vec{t} \cdot \nabla \mv{A} = \left[ \frac{\dd \mv{A}}{\dd \tau} + \frac{\dd \inv{\Lambda}}{\dd \tau} (\mv{A}) \right]_{\tau = 0}.
\eea
The type of transformation that does not affect the ``absolute value'' of a multivector is a (proper, orthochronous) Lorentz transformation, which we write as
\bea
\mv{A}' = \op{\Lambda}(\mv{A}) = \tilde{\mv{U}} \mv{A} \mv{U} \quad \text{where} \quad \tilde{\mv{U}}\mv{U} = 1
\eea
so that $\mv{A}'{}^2 = \mv{A}^2$ for any multivector $\mv{A}$.  $\mv{U}$ is an even multivector called a \emph{rotor}, which we write as
\bea
\mv{U} = \text{e}^{\mv{B}/2} \qquad \mv{B} = - \tilde{\mv{B}},
\eea
so that $\mv{B}$ is a bivector.  If we let $\mv{B}(\tau = 0) = 0$, we find that
\bea
\left.\frac{\dd \inv{\Lambda}}{\dd \tau} (\mv{A})\right|_{\tau = 0} = \left[ \frac{\dd \mv{U}}{\dd \tau} \mv{A}\tilde{\mv{U}} + \mv{UA} \frac{\dd \tilde{\mv{U}}}{\dd \tau} \right]_{\tau = 0} = \frac{1}{2} \left[ \frac{\dd\mv{B}}{\dd \tau} \mv{A} - \mv{A} \frac{\dd\mv{B}}{\dd \tau} \right]_{\tau = 0} = \left. \frac{\dd\mv{B}}{\dd \tau}\right|_{\tau = 0} \times \mv{A},
\eea
where ``$\times$'' represents the commutator product $(\mv{A}\mv{B}-\mv{B}\mv{A})/2$.  Thus, if we let
\bea
\vec{t} \cdot \hat{\nabla} \mv{A} \equiv \left.\frac{\dd\mv{A}}{\dd\tau}\right|_{\tau=0} \qquad \text{and} \qquad \tensor{\Gamma}(\vec{t},\mv{A}) \equiv \left. \frac{\dd\mv{B}}{\dd \tau}\right|_{\tau = 0} \times \mv{A} \equiv \mv{\Omega}(\vec{t}) \times \mv{A},
\label{biv_connection}
\eea
we have the covariant derivative in the form of \eq{cov_decomp}, with $\mv{\Omega}$ being the same bivector connection as in \eq{cov_deriv}.  The connection bivector maps a vector, which is associated with a path in spacetime, onto a bivector, which is related to the (four-dimensional) rotation of the orthonormal frame as it is carried from one point in the manifold to another.

\subsection{Curvature Tensors and Einstein's Equations}

In general, covariant derivatives will not commute:
\bea
\left[ \vec{a} \cdot \acute{\nabla}, \vec{b} \cdot \acute{\nabla}  \right] 
\acute{\mv{A}} \equiv  
\left( \vec{a} \cdot \nabla \vec{b} \cdot \nabla 
- \vec{b} \cdot \nabla \vec{a} \cdot \nabla \right) \mv{A} 
- \left[ \vec{a}, \vec{b} \right] \cdot \nabla \mv{A} 
\label{curvature_equation}
\eea
where 
\bea
\left[ \vec{a}, \vec{b} \right] \equiv 
\vec{a}\cdot \nabla \vec{b} - \vec{b}\cdot \nabla\vec{a}
\eea
denotes the Lie bracket (we are assuming throughout that torsion vanishes). The accent marks in the first expression
show that the derivatives are to act only on the multivector, not
on the vectors $\vec{a}$ and $\vec{b}$. Using \eq{cov_deriv} for
the covariant derivatives, we define the Riemann curvature tensor
$\tensor{R}$ as
\bea
\left[ \vec{a} \cdot \acute{\nabla}, \vec{b} \cdot \acute{\nabla}  \right] 
\acute{\mv{A}} &=& \tensor{R}(\vec{a}\wedge\vec{b})\times\mv{A},\nonumber \\
\tensor{R}(\vec{a} \wedge \vec{b})&\equiv& \vec{a}\cdot \hat{\nabla} \mv{\Omega}(\vec{b}) 
- \vec{b}\cdot \hat{\nabla} \mv{\Omega}(\vec{a}) 
+ \mv{\Omega}(\vec{a})\times\mv{\Omega}(\vec{b}) - \mv{\Omega}\left( \left[ \vec{a}, \vec{b} \right]\right). 
\label{riemann_def}
\eea
This object is a bivector-valued function of bivectors which is
linear in both of its arguments, due to the linearity of the derivative
operators. Also, the outer product guarantees that both sides vanish
if $\vec{a}$ and $\vec{b}$ are not linearly independent.
Geometrically, the grade of the curvature tensor and its argument
are sensible: the argument specifies a particular spacetime plane, and
if a multivector is parallel-transported around an infinitesimal closed path
contained in this plane, it will undergo a rotation described by
the bivector returned by the Riemann tensor.

The Riemann tensor obeys the Ricci identity (see \cite{SCGT}, section 11 for derivations)
\begin{subequations}
\bea
  &{}& \vec{\partial_a} \wedge \tensor{R}(\vec{a}\wedge\vec{b}) = 0 \\
  &{}& \vec{a} \cdot \tensor{R}(\vec{b}\wedge\vec{c}) + \vec{b} \cdot \tensor{R}(\vec{c}\wedge\vec{a}) + \vec{c} \cdot \tensor{R}(\vec{a}\wedge\vec{b}) = 0 \label{ricci_ident},
\eea
\end{subequations}
the symmetry condition
\bea
	(\vec{a}\wedge\vec{b}) \cdot \tensor{R}(\vec{c}\wedge\vec{d}) = \tensor{R}(\vec{a}\wedge\vec{b}) \cdot (\vec{c}\wedge\vec{d}),
\eea
and the Bianchi identity
\begin{subequations}
\bea
&{}& \acute{\nabla}\wedge \acute{\tensor{R}}(\vec{a}\wedge\vec{b}) = 0 
\label{bianchi_curl}\\
	&{}& \vec{a}\cdot \nabla \tensor{R}(\vec{b}\wedge\vec{c}) 
+ \vec{b}\cdot \nabla \tensor{R}(\vec{c}\wedge\vec{a}) 
+ \vec{c}\cdot \nabla \tensor{R}(\vec{a}\wedge\vec{b}) = 0. 
\label{bianchi_ident}
\eea
\end{subequations}
In both the Ricci and Bianchi identities, the ``a'' and ``b'' expressions 
are equivalent, although \eq{ricci_ident} and \eq{bianchi_ident} are the 
simpler ones to use in calculations.  

Contraction of the Riemann tensor yields the Ricci tensor:
\bea
\tensor{R}(\vec{b}) = \vec{\partial_a} \cdot \tensor{R}(\vec{a}\wedge\vec{b}),
\eea
which is vector-valued and has the symmetry properties
\bea
\vec{\partial_a}\wedge \tensor{R}(\vec{a}) = 0
\eea
and
\bea
\vec{a}\cdot\tensor{R}(\vec{b}) = \tensor{R}(\vec{a})\cdot\vec{b}.
\eea
One last contraction yields the Ricci scalar:
\bea
R = \vec{\partial_a}\cdot \tensor{R}(\vec{a}).
\eea
The Einstein tensor is defined in terms of the Ricci tensor and scalar as
\bea
\tensor{G}(\vec{a}) \equiv \tensor{R}(\vec{a}) - \frac{1}{2} \vec{a} R.
\eea
The Bianchi identity, \eq{bianchi_curl},
can then be rewritten in an illustrative form by
contracting out the vectors $\vec{a}$ and $\vec{b}$:
\bea
\vec{\partial_a} \cdot \left[ \vec{\partial_b} \cdot 
\left( \acute{\nabla}\wedge \acute{\tensor{R}}(\vec{a}\wedge\vec{b}) 
\right) \right] = \acute{\tensor{R}}(\acute{\nabla}) - \frac{1}{2} \nabla R = 0
\eea
or, more compactly,
\bea
\acute{\tensor{G}} (\acute{\nabla}) = 0.
\eea

Finally, the Weyl tensor $\tensor{C}(\vec{a}\wedge\vec{b})$ is defined by 
\bea
\tensor{C}(\vec{a}\wedge\vec{b})\equiv \tensor{R}(\vec{a}\wedge\vec{b})
- \frac{1}{2}\left(\tensor{R}(\vec{a})\wedge\vec{b} 
+ \vec{a}\wedge\tensor{R}(\vec{b})\right) 
+ \frac{1}{6}\vec{a}\wedge\vec{b}\tensor{R}.
\label{weyl_def}
\eea
$\tensor{C}$ is antisymmetric in its arguments, and tractionless:
\bea
\tensor{C}(\vec{a}\wedge \vec{b}) = - \tensor{C}(\vec{b}\wedge\vec{a}) \qquad \vec{\partial_a} \tensor{C}(\vec{a}\wedge\vec{b}) = 0.
\eea
Section~\ref{petrov_class} is devoted to properties of this tensor.

%\subsection{The Stress-Energy Tensor and Einstein Equations}

The Einstein field equations are
\bea
\tensor{G}(\vec{a}) = 8 \pi \tensor{T}(\vec{a})
\eea
in units with $G=1$, where vector-valued $\tensor{T}(\vec{a})$ 
is the stress-energy tensor. For a perfect fluid with density $\rho$ and
pressure $p$, the stress-energy tensor is \cite{GTG}
\bea
\tensor{T}^{(EM)}(\vec{a}) = (\rho + P) \vec{a} \cdot \vec{u} \vec{u} - P \vec{a},
\eea
where $\rho$ is the density, $P$ is the pressure, and $\vec{u}^2 = 1$ is the 4-velocity of an observer comoving with the fluid.  Another useful form is for electromagnetic fields with field
strength bivector $\mv{F}$; the stress-energy tensor is then \cite{riesz,STA}
\bea
\tensor{T}^{(EM)}(\vec{a}) = - \frac{1}{2} \mv{F} \vec{a} \mv{F}.
\eea

\subsection{The Coordinate Frame}
\label{frames}

After some work, we find the curl of a reciprocal frame vector is
\bea
\nabla \wedge \vec{e}^\mu = \frac{1}{2} (\vec{e}^\alpha \wedge \vec{e}^\beta)
\left( \left[ \vec{e}_\alpha, \vec{e}_\beta \right] \cdot \vec{e}^\mu \right),
\eea
which is the Maurer-Cartan equation in curved spacetime \cite{nakahara}.
A particularly useful type of frame is
the coordinate frame $\{\vec{g}^\mu\}$ defined (in the absence of torsion)
by the holonomic condition
\bea
\left[ \vec{g}_\mu, \vec{g}_\nu \right] = 0 
\qquad \text{or} \qquad \vec{g}_\mu \cdot \nabla \vec{g}_\nu = 
\vec{g}_\nu \cdot \nabla \vec{g}_\mu,
\label{holonomic_frame}
\eea
which leads immediately to
\bea
\nabla \wedge \vec{g}^\mu \equiv 
\vec{\partial_a} \wedge \left( \vec{a} \cdot \nabla \vec{g}^\mu \right) 
= \hat{\nabla} \wedge \vec{g}^\mu + \vec{\partial_a} \wedge 
\left( \mv{\Omega}(\vec{a}) \cdot \vec{g}^\mu \right) = 0 .
\label{torsion_free}
\eea  
This Cartan's first structural equation in a disguised form; 
see Section \ref{diffgeom} for an explicit translation of the differential 
form language.  Unless otherwise noted, vector and multivector
components will be written  in terms of the coordinate frame; for example,
\beann
b_\mu = \vec{b} \cdot \vec{g}_\mu \ , \quad 
\nabla_\mu = \vec{g}_\mu \cdot \nabla, \quad\text{etc.}
\eeann

As shown in Appendix~\ref{connection_derivation}, we can 
invert \eq{torsion_free} to obtain the connection:
of a coordinate frame vector \cite{snygg}:
\begin{subequations}
\bea
\mv{\Omega}(\vec{b}) &=& \frac{1}{2} \left( \vec{g}^\mu \wedge \hat{\nabla} b_\mu - \vec{g}_\mu \wedge \hat{\nabla} b^\mu + \vec{g}^\mu \wedge \left( \vec{b}\cdot \hat{\nabla} \vec{g}_\mu \right) \right) \\
\mv{\Omega}_\alpha &\equiv& \mv{\Omega}(\vec{g}_\alpha ) = 
\frac{1}{2} \left\{ \vec{g}^\nu \wedge \hat{\nabla} (\vec{g}_\alpha \cdot \vec{g}_\nu ) 
+ \vec{g}^\mu \wedge \hat{\nabla}_\alpha \vec{g}_\mu \right\} ,
\label{connection_solution}
\eea
\end{subequations}
where $\hat{\nabla}_\mu \equiv \vec{g}_\mu \cdot \hat{\nabla}$; the second expression is for the specific case of a frame vector, which is often the more useful form.  We may write the Riemann tensor over frame vectors as
\bea
\tensor{R}_{\mu\nu} = \tensor{R}(\vec{g}_\mu \wedge\vec{g}_\nu) 
= \vec{g}_\mu \cdot \hat{\nabla} \mv{\Omega}_\nu 
- \vec{g}_\nu \cdot \hat{\nabla} \mv{\Omega}_\mu 
+ \mv{\Omega}_\mu \times \mv{\Omega}_\nu,
\label{riemann_tensor}
\eea
while the Ricci tensor is
\bea
\tensor{R}_\mu = \tensor{R}(\vec{g}_\mu) 
= \vec{\partial_a} \cdot \tensor{R}(\vec{a}\wedge\vec{g}_\mu ) 
= \vec{g}^\nu \cdot \tensor{R}_{\nu\mu}.
\label{ricci_tensor}
\eea
The usual (scalar) components of these tensors are, respectively,
\beann
R^{\kappa}_{\lambda\mu\nu} = (\vec{g}^\kappa \wedge \vec{g}_\lambda)\cdot \tensor{R}_{\mu\nu} \qquad R_{\mu\nu} = \vec{g}_\mu \cdot \tensor{R}_\nu ;
\eeann
the same process can be used to obtain components of any tensor object, such as the stress-energy tensor of a perfect fluid:
\beann
T^{(PF)}_{\mu\nu} = \vec{g}_\mu \cdot \tensor{T}^{(PF)}(\vec{g}_\nu) = (\rho + P)u_\mu u_\nu - P \vec{g}_\mu \cdot \vec{g}_\nu ,
\eeann
where $u_\nu = \vec{u} \cdot \vec{g}_\nu$.

Finally, we can assign a linear operator to map
the orthonormal frame into the coordinate frame: 
\bea 
\vec{g}_\mu =
\inv{h}(\hvec{e}_\mu) = h^a_\mu \hvec{e}_a.
\eea 
This operator is a representation of the
vierbein field; it occupies a prominent position 
in the gauge theory formulation
of gravity given by Lasenby, Doran, and Gull \cite{GTG}, 
where it acts as a frame map,
lifting a frame of vectors in flat spacetime to a generally-covariant
spacetime \cite{SCGT}.  The coordinate frame also defines the
\emph{metric components} via 
\bea 
\vec{g}_\mu \cdot \vec{g}_\nu \equiv
g_{\mu\nu} .
\eea
%and the metric compatibility condition is 
%\bea
%\vec{a} \cdot \nabla g_{\mu\nu} = \vec{a} \cdot \pd g_{\mu\nu} = \tensor{\Gamma}(\vec{a},\vec{g}_\mu ) \cdot \vec{g}_\nu + \tensor{\Gamma}(\vec{a},\vec{g}_\nu) \cdot \vec{g}_\mu
%\label{metric_compat}
%\eea
%where $\tensor{\Gamma}(\vec{a},\vec{g}_\mu) \equiv \vec{a} \cdot \nabla \vec{g}_\mu$ is the Christoffel connection vector, whose components are
%\bea
%\Gamma^{\lambda}_{\mu\nu} = \Gamma^{\lambda}_{\nu\mu} = \vec{g}^\lambda \cdot \tensor{\Gamma}(\vec{g}_\mu,\vec{g}_\nu).
%\eea
%Please note that the metric components are treated as scalars, rather than as a tensor, so that \eq{metric_compat} looks trivial.  However, the symmetry of the Christoffel symbols is the result of our choice of covariant derivative.
We emphasize that in the formulation outlined here, the metric can be seen as a by-product of the choice of some particular coordinate frame. 
All of the geometrical content of gravitation is defined without
reference to the spacetime metric $g_{\mu\nu}$. In order to define the
notion of parallel transport which preserves lengths of vectors, we
have alternatively assumed the existence of an orthonormal frame of
vectors at each point in the manifold, in contrast to most textbook treatments
of general relativity, in which the metric is the central mathematical
object (despite its lack of direct physical content). However, it should be pointed out that both starting points give mathematically equivalent descriptions.

\section{A Method for Direct Calculation and the Schwarzschild Solution}

\subsection{General Solution Technique}
\label{lexicon}

Given the constructions of the previous section, we have an explicit expression for the connection bivectors $\mv{\Omega}_\nu$ in terms of the coordinate basis vectors and the curvature bivectors $\tensor{R}_{\mu\nu}$ in terms of the connection; thus the Einstein equations can be viewed as differential equations in the coordinate frame vectors.  Symmetries of the system can be incorporated directly into the definition of the coordinate vector.

For a given physical system which has well-defined symmetries,
\begin{enumerate}
	\item \label{frame_symmetry} Define an orthonormal frame $\{\hvec{e}_\mu\}$ and its reciprocal $\{\hvec{e}^\mu\}$ that corresponds to the symmetry at hand (\emph{e.g.}, spherical coordinates for spherical symmetry, axisymmetric coordinates if a preferred axis exists, and so on)
	\item \label{physical_frame} Define a frame $\vec{g}_\mu = \inv{h}(\hvec{e}_\mu)$ that preserves the symmetry of item \ref{frame_symmetry};
	\item Define the derivative to preserve the orthonormal frame:  $\vec{a} \cdot \hat{\nabla} \hvec{e}_\mu = 0$, and decompose the covariant derivative as $\vec{a}\cdot \nabla \mv{A} = \vec{a}\cdot\hat{\nabla} + \mv{\Omega}(\vec{a})\times\mv{A}$;
	\item Calculate the connection bivectors, using \eq{connection_solution} \[ \mv{\Omega}_\alpha \equiv \mv{\Omega}(\vec{g}_\alpha ) = \frac{1}{2} \left( \vec{g}^\nu \wedge \hat{\nabla} g_{\alpha \nu} + \vec{g}^\mu \wedge \hat{\nabla}_\alpha \vec{g}_\mu \right)	;\]
	\item Calculate the Riemann curvature bivectors:  \[\tensor{R}_{\mu\nu} = \vec{g}_\mu \cdot \hat{\nabla} \mv{\Omega}_\nu - \vec{g}_\nu \cdot \hat{\nabla} \mv{\Omega}_\mu + \mv{\Omega}_\mu \times \mv{\Omega}_\nu ; \]
	\item Find the Ricci vectors: \[ \tensor{R}_\mu = \vec{g}^\nu \cdot \tensor{R}_{\nu\mu};\]
	\item \label{special_case} Solve the vacuum equations for the coordinate frame vectors $\{\vec{g}_\nu\}$:  \[ \tensor{R}_\mu = 0. \] or the field equations in the presence of some source of energy-momentum: \[ \tensor{R}_\mu = 8 \pi \left( \tensor{T}(\vec{g}_\mu) - \frac{1}{2} \vec{g}_\mu T \right), \] where $T = \vec{g}^\mu \cdot \tensor{T}(\vec{g}_\mu)$.
\end{enumerate}

% Fix this!!!
We propose this technique as a replacement for the usual ``brute force'' component method, as well as providing an alternative to the method of differential forms in cases of special symmetries.  As we will see in our discussion of the Kerr solution in Section~\ref{kerrsol}, we often do not even need to specify a particular coordinate system to extract a great deal of information from the connection and curvature tensors.  This method combines the efficiency of forms with the straightforwardness of coordinate methods, and as such we believe can fill the role of either.

\subsection{Example:  Schwarzschild Geometry}
\label{schwarz_geom}

We now use the recipe laid out in the previous section to obtain the exact solution to Einstein's vacuum equations for the case of stationary, spherically-symmetric spacetimes.  The resulting black hole solution is of course equivalent to Schwarzshild's solution, but we provide a more general solution, which in a single expression contains a large class of equivalent metric forms, of which the Schwarzshild metric is a special choice.  In Section~\ref{kerrsol} we will examine the more complicated case of the rotating black hole, after we examine techniques to classify and study spacetimes.

\subsubsection{Conditions for Spherical Symmetry}

Flat spacetime is trivially invariant under spatial rotations; in fact, every point may be considered the center of a sphere for the purposes of rotations.  For the non-trivial vacuum case in curved spacetime, we will pick a single point as an origin in the orthonormal frame and write a set of basis vectors in spherical coordinates:
\bea
	\hvec{e}_\mu \cdot \hvec{e}_\nu = \eta_{\mu\nu}
\eea
where $\hvec{e}_\mu = \hvec{e}_\mu (\theta,\phi)$ in general, for the angular variables $\theta$ and $\phi$.  We define a rotation as
\bea
	\op{R}(\mv{A}) = \mathrm{e}^{-\frac{1}{2}\mv{B}} \mv{A} \mathrm{e}^{\frac{1}{2}\mv{B}}
\eea
where $\mv{B}=\mv{B}(\vec{x})$ is a spacelike bivector.  The timelike direction is invariant under rotations:  $\mv{B}\cdot \hvec{e}_0 = 0$.  Our ``spherical coordinates'' reflect this condition in the orthonormal frame.

The vierbein fields should preserve this spherical symmetry, while allowing for distortion in the radial and temporal directions.  In other words, we want the rotation operator to commute with the vierbein operation:
\bea
	\inv{h}\left( \op{R}\mv{A}\right) = \op{R} \inv{h}(\mv{A}).
\eea
Therefore, define a set of orthogonal basis vectors in spherical coordinates $\vec{e}^\mu = \vec{\partial}x^\mu$ for the orthonormal frame:
\bea
\begin{array}{lll}
\vec{e}^0 = \vec{\gamma}^0 &\quad & \vec{e}^1 = \sin \theta \cos \phi \vec{\gamma}^1 + \sin \theta \sin \phi \vec{\gamma}^2 + \cos \theta \vec{\gamma}^3 \\
\vec{e}^2 = r^{-1} \left(\cos \theta \cos \phi \vec{\gamma}^1 + \cos \theta \sin \phi \vec{\gamma}^2 - \sin \theta \vec{\gamma}^3 \right) & \quad & \vec{e}^3 = (r \sin \theta)^{-1} \left( - \sin \phi \vec{\gamma}^1 + \cos \phi \vec{\gamma}^2 \right)
\end{array}
\eea
where the $\{\vec{\gamma}^\mu\}$ are the ordinary Cartesian basis vectors for four dimensions, with $(\vec{\gamma}^0)^2 =1$ and $(\vec{\gamma}^i)^2 = -1$ for $i = 1,2,3$.  We normalize these to make their inner product constant:
\bea
\hvec{e}^2 = r \vec{e}^2 \quad \text{and} \quad \hvec{e}^3 = r \sin \theta \vec{e}^3,
\eea
while $\hvec{e}^0 = \vec{e}^0$ and $\hvec{e}^1 = \vec{e}^1$.  Now $\hvec{e}^\mu \cdot \hvec{e}^\nu = \eta^{\mu\nu}$, where $\eta = \mathrm{diag} (1,-1,-1,-1)$ is the metric.

The $\vec{e}^0$ and $\vec{e}^1$ directions are invariant under any rotation, while $\vec{e}^2$ and $\vec{e}^3$ are rotated into each other; any set of basis vectors of this type are spherically-symmetric.  In particular, a set of basis vectors in the gauge-covariant space are automatically spherically-symmetric if they take the following form \cite{GTG}:
\bea
\begin{array}{lll}
        \vec{g}^t = \adj{h}(\hvec{e}^0) = f_1 \hvec{e}^0 + f_2 \hvec{e}^1 & \quad & \vec{g}^r = \adj{h}(\hvec{e}^1) = g_1 \hvec{e}^1 + g_2 \hvec{e}^0 \\
	\vec{g}^\theta = \adj{h}(\hvec{e}^2) = r^{-1} (\alpha \hvec{e}^2 + \beta \hvec{e}^3 ) & \quad & \vec{g}^\phi = \adj{h}(\hvec{e}^3) = (r \sin \theta)^{-1} ( \alpha \hvec{e}^3 - \beta \hvec{e}^2 )
\end{array}
\label{sphere_basis}
\eea
where the $f_i$, $g_i$, $\alpha$, and $\beta$ are functions of $r$ and $t$.  Note, however, that there is no loss in generality in letting $\beta = 0$ and $\alpha = 1$, since we can effectively perform this operation via rotations and judicious choice of the radial variable $r$. 

The stationary condition amounts ultimately to the independence of all quantities from the time variable $t$ \cite{wald}.  Note that this is \emph{not} the same as the static condition, which is a statement that $\vec{g}^t \cdot \vec{g}^r = 0$.  The static condition is highly restrictive, and leads automatically to the Schwarzschild metric.  The stationary condition is sufficient to derive a general expression for spherically-symmetric spacetimes (though we need not even assume that---see \emph{e.g.} \cite{carroll}), as we now show.

\subsubsection{Solution of the Field Equations}

Applying the conditions $\alpha =1$ and $\beta=0$ to \eq{sphere_basis} yields the following co- and contravariant basis vectors:
\bea
\begin{array}{lll}
	\vec{g}^t = \adj{h}(\hvec{e}^0) = f_1 \hvec{e}^0 + f_2 \hvec{e}^1 & \quad & 
	\vec{g}_t = \inv{h}(\hvec{e}_0) = g_1 \hvec{e}_0 - g_2 \hvec{e}_1 \\
	\vec{g}^r = \adj{h}(\hvec{e}^1) = g_1 \hvec{e}^1 + g_2 \hvec{e}^0 & \quad & 
	\vec{g}_r = \inv{h}(\hvec{e}_1) = f_1 \hvec{e}_1 - f_2 \hvec{e}_0 \\
	\vec{g}^\theta = \adj{h}(\hvec{e}^2) = r^{-1} \hvec{e}^2 & \quad & 
	\vec{g}_\theta = \inv{h}(\hvec{e}_2) = r \hvec{e}_2 \\
	\vec{g}^\phi = \adj{h}(\hvec{e}^3) = (r \sin \theta)^{-1} \hvec{e}^3 & \quad & 
	\vec{g}_\phi = \inv{h}(\hvec{e}_3)  = r \sin \theta \hvec{e}_3
\end{array}
\label{basis_vectors}
\eea
where the functions $f_i$ and $g_i$ are assumed to depend on $r$ only, and we have imposed a simple gauge condition
\bea
	f_1 g_1 - f_2 g_2 = 1 .
\label{gauge_condition}
\eea
This gauge condition is actually the only one we need to solve the vacuum equations; it guarantees that the frames in \eq{basis_vectors} satisfy the torsion-free condition in \eq{torsion_free}.

The directional derivatives of the basis vectors are
\bea
\begin{array}{lll}
	\vec{g}_t \cdot \hat{\nabla} \vec{g}_\nu = 0 & \quad &
	\vec{g}_\phi \cdot \hat{\nabla} \vec{g}_\nu = 0 \\
	\vec{g}_r \cdot \hat{\nabla} \vec{g}_t = g'_1 \hvec{e}_0 - g'_2 \hvec{e}_1 & \quad &
	\vec{g}_r \cdot \hat{\nabla} \vec{g}_\theta = \hvec{e}_2 \\
	\vec{g}_r \cdot \hat{\nabla} \vec{g}_r = f'_1 \hvec{e}_1 - f'_2 \hvec{e}_0 & \quad &
	\vec{g}_r \cdot \hat{\nabla} \vec{g}_\phi = \sin \theta \hvec{e}_3 \\
	\vec{g}_\theta \cdot \hat{\nabla} \vec{g}_\phi = r \cos \theta \hvec{e}_3 & \quad &
	\vec{g}_\theta \cdot \hat{\nabla} \vec{g}_{\nu \neq \phi} = 0
\end{array}
\eea
where the primes indicate differentiation with respect to $r$.  Plugging these into \eq{connection_solution}, we obtain the connection bivectors (after some work):
\bea
	\mv{\Omega}_t &=& - \left( g_1 g'_1 - g_2 g'_2 \right) \hvec{e}_0\hvec{e}_1 = - \frac{1}{2} \partial_r (g^2_1 - g^2_2) \hvec{e}_0\hvec{e}_1 \nonumber \\
	\mv{\Omega}_r &=& \left( g'_1 f_2 - g'_2 f_1 \right) \hvec{e}_0\hvec{e}_1 \nonumber \\
	\mv{\Omega}_\theta &=& g_1 \hvec{e}_1\hvec{e}_2 - g_2 \hvec{e}_0\hvec{e}_2 \nonumber \\
	\mv{\Omega}_\phi &=& \sin\theta \left( g_1 \hvec{e}_1\hvec{e}_3 - g_2 \hvec{e}_0\hvec{e}_3 \right) + \cos\theta \hvec{e}_2\hvec{e}_3
\label{schwarzconnect}
\eea
where we have used the fact that $\hvec{e}_\mu \wedge \hvec{e}_\nu = \hvec{e}_\mu \hvec{e}_\nu$ when $\mu \neq \nu$.  Note that no derivatives of the $f_i$ functions appear.

The six Riemann curvature bivectors follow from \eq{riemann_tensor} and take the compact forms
\bea
	\tensor{R}_{r t} &=& \vec{g}_r \cdot \hat{\nabla} \mv{\Omega}_t = -\frac{1}{2} \partial^2_r \left( g^2_1 - g^2_2 \right) \vec{g}_t \wedge \vec{g}_r \nonumber \\
	\tensor{R}_{\theta t} &=& \mv{\Omega}_\theta \times \mv{\Omega}_t = \frac{1}{2} \partial_r (g^2_1 - g^2_2) \frac{\vec{g}_\theta \wedge \vec{g}_t}{r} \nonumber \\
	\tensor{R}_{\phi t} &=& \mv{\Omega}_\phi \times \mv{\Omega}_t = \frac{1}{2} \partial_r (g^2_1 - g^2_2) \frac{\vec{g}_\phi \wedge \vec{g}_t}{r} \nonumber \\
	\tensor{R}_{\theta r} &=& - \vec{g}_r \cdot \hat{\nabla} \mv{\Omega}_\theta + \mv{\Omega}_\theta \times \mv{\Omega}_r = \frac{1}{2} \partial_r (g^2_1 - g^2_2) \frac{\vec{g}_\theta \wedge \vec{g}_r}{r} \nonumber \\
	\tensor{R}_{\phi r} &=& - \vec{g}_r \cdot \hat{\nabla} \mv{\Omega}_\phi + \mv{\Omega}_\phi \times \mv{\Omega}_r = \frac{1}{2} \partial_r (g^2_1 - g^2_2) \frac{\vec{g}_\phi \wedge \vec{g}_r}{r} \nonumber \\
	\tensor{R}_{\phi \theta} &=& - \vec{g}_\theta \cdot \hat{\nabla} \mv{\Omega}_\phi + \mv{\Omega}_\phi \times \mv{\Omega}_\theta = \left( (g^2_1 - g^2_2) - 1 \right) \frac{\vec{g}_\phi \wedge \vec{g}_\theta}{r^2} .
\eea
Note that all of these equations contain the quantity $g^2_1 - g^2_2$ (which is the $g_{00}$ term of the metric).  The Ricci curvature tensor (via \eq{ricci_tensor}) is particularly simple:
\bea
	\tensor{R}_t &=& \Phi(r) ( g_1 \hvec{e}_0 - g_2 \hvec{e}_1 ) = \Phi(r) \vec{g}_t \nonumber \\
	\tensor{R}_r &=& \Phi(r) ( f_1 \hvec{e}_1 - f_2 \hvec{e}_0 ) = \Phi(r) \vec{g}_r \nonumber \\
	\tensor{R}_\theta &=& \Psi(r) \hvec{e}_2 = (\Psi(r)/r) \vec{g}_\theta \nonumber \\
	\tensor{R}_\phi &=& \Psi(r) \sin\theta \hvec{e}_3 = (\Psi(r)/r) \vec{g}_\phi
\label{ricci_values}
\eea
where
\begin{subequations}
\bea
	\Phi(r) &\equiv& \frac{1}{2} \frac{\partial^2}{\partial r^2} (g^2_1 - g^2_2) + \frac{1}{r} \frac{\partial}{\partial r} (g^2_1 - g^2_2) = \frac{1}{2 r} \frac{\partial^2}{\partial r^2} \left( r (g^2_1 - g^2_2) \right) \\
	\Psi(r) &\equiv& \frac{\partial}{\partial r} (g^2_1 - g^2_2) + \frac{1}{r} ( g^2_1 - g^2_2 - 1) = \frac{1}{r} \left\{ \frac{\partial}{\partial r} \left( r (g^2_1 - g^2_2) \right) - 1 \right\}.
\eea
\end{subequations}
In the vacuum case, $\tensor{R}_\mu = 0$ identically, reducing in our case to $\Phi(r)=0$ and $\Psi(r)=0$, which are differential equations in the quantity $g^2_1 - g^2_2$.  Letting $\chi(r) = g^2_1 - g^2_2$, the differential equations become
\bea
	\frac{\partial^2}{\partial r^2} \left( r \chi(r) \right) = 0 \quad \text{and} \quad
	\frac{\partial}{\partial r} \left( r \chi(r) \right) - 1 = 0
\eea
with the solution
\bea
	\chi(r) \equiv g^2_1 - g^2_2 = 1 - \frac{K}{r}
\label{solution}
\eea
where $K$ is a constant of integration.  From past knowledge of the Schwarzschild metric, we know that $K = 2 M$.

Note that the solution in \eq{solution} and the gauge condition from \eq{gauge_condition} are the only restrictions the vacuum equations can give us.  We cannot specify from the preceding analysis alone what \emph{any} of the quantities $f_i$ and $g_i$ are by themselves.  Thus, the most general line element satisfying the field equations is
\begin{subequations}
\bea
	\dd\tau^2 &=& \vec{g}_\mu \cdot \vec{g}_\nu \dd x^\mu \dd x^\nu = \chi(r) \dd t^2 + 2 (f_1 g_2 - f_2 g_1) \dd t \dd r - (f^2_1 - f^2_2 ) \dd r^2 - r^2 \dd \Omega^2 \\
	\chi(r) &=& 1 - \frac{2 M}{r} \\
	f_1 g_1 &-& f_2 g_2 = 1
\eea
\label{line_element}
\end{subequations}
We note that \eq{line_element} specifies the ``radial'' piece $g^{11}$ of the inverse metric only, since the ``coordinates'' are related to the covariant basis elements $\vec{g}^\mu$.  Thus, the time part of the gauge is completely free to be chosen, including its direction relative to the radial part.  This freedom allows for the veritable zoo of forms known for the vacuum metric, which can all be obtained from each other via coordinate transformations due to Birkhoff's theorem.  A closer examination of \eq{line_element} shows that there is only one free function of radius, which is the combination $B(r) = f_1 g_2 - f_2 g_1$; once this is determined, the other functions display all the possible isometries.

Particular choices of the $f_i$ and $g_i$ yield familiar forms for the metric. When $f_1 g_2 - f_2 g_1= 0$, the ordinary Schwarzschild metric is obtained,  while $f^2_1 - f^2_2 = 0$ yields the Eddington-Finkelstein metric (advanced and retarded, depending on other sign choices), and $f^2_1 - f^2_2 = 1$ gives us the Painlev\'{e}-Gullstrand line-element, discovered independently by Lasenby, Doran, and Gull \cite{GTG}.  For the class of solutions $f^2_1 - f^2_2 = const.$, which includes the latter two, see \cite{martel} and the references therein.  The Kruskal extension (see \emph{e.g.} \cite{wald}) requires assigning two separate vierbein fields $\op{h}$ to each point in spacetime, which provides the double coordinate cover.  Since this procedure lies somewhat outside the scope of this paper, we will not discuss it further here.

It should be noted that \eq{line_element} can be obtained via the usual method of differential forms \cite{zapolsky}, although the authors are not aware of any publication of this result.  This fact is hardly surprising, since there is not any essentially new information contained in this line element.  However, we would like to emphasize that the method of geometric algebra is completely turn-crank, and, unlike the method of differential forms, the connection coefficients can be found directly, as opposed to solving algebraic equations (see \cite{compgrav}, for example).  In fact, the solution shown here was obtained not only by hand calculations, but via a short program written by one of the authors \footnote{This program was written by M. R. Francis in Maple (\texttt{http://www.maplesoft.com/}), using Mark Ashdown's GA package, which can be downloaded at \texttt{http://www.mrao.cam.ac.uk/\~{}clifford/software/GA/}.  The complete calculation on the author's home computer took just over one minute.}.

\section{Translation of Differential Forms}
\label{diffgeom}

\subsection{The Exterior Derivative in Geometric Algebra}

For the sake of comparison with the standard approaches to general relativity (see \cite{wald,misner}), it is useful to understand how the geometric algebra formalism relates to the differential-geometric formalism.  To do this, we will proceed to incorporate the language of differential forms into geometric algebra wholesale.

Before doing anything else, we explicitly identify differential forms with pure multivectors (blades).  (Hestenes and Sobczyk use a slightly different convention, wherein differential forms are scalar-valued \cite{CA2GC}; see also \cite{GAdiff}.)  We may then expand an arbitrary $r$-blade in the coordinate basis:
\bea
\mv{\omega} = \frac{1}{r!} \omega_{a_1 a_2 \ldots a_r} \DD x^{a_1} \wedge \DD x^{a_2} \wedge \ldots \wedge \DD x^{a_r}
\eea
where $\DD$ is the exterior differential, which we need to understand in the language of geometric algebra.  To do so, we take the derivative of our arbitrary $r$-blade, and obtain an $r+1$-blade:
\bea
\DD \mv{\omega} = \frac{1}{r!} \left( \frac{\partial \omega_{a_1 a_2 \ldots a_r}}{\partial x^b} \right) \DD x^b \wedge \DD x^{a_1} \wedge \DD x^{a_2} \wedge \ldots \wedge \DD x^{a_r}.
\label{extdiff}
\eea
Obviously, $\DD x^b$ is a vector (one-form), while $x^b$ is a scalar function.  This indicates that $\DD$ is vector-valued, and acts via the exterior product (as expected from its name).  Applying the operator a second time yields
\bea
\DD^2 \mv{\omega} = \frac{1}{r!} \left( \frac{\partial^2 \omega_{a_1 a_2 \ldots a_r}}{\partial x^c \partial x^b} \right) \DD x^c \wedge \DD x^b \wedge \DD x^{a_1} \wedge \DD x^{a_2} \wedge \ldots \wedge \DD x^{a_r} = 0
\eea
since the action of partial derivatives commutes, while the wedge product changes sign when two adjacent arguments are interchanged.

Since the coordinate functions $x^a$ are scalars, the action of the exterior differential is the same as the gradient is the same as the covariant gradient or curl:
\bea
\DD x^a = \vec{\partial} x^a = \nabla x^a = \nabla \wedge x^a = \vec{g}^a.
\eea
For generality's sake (and following the precedent set by such texts as \cite{misner}), we will consider the covariant gradient.  In this case,
\bea
\mv{\omega} = \frac{1}{r!} \omega_{a_1 a_2 \ldots a_r} \nabla x^{a_1} \wedge \nabla x^{a_2} \wedge \ldots \wedge \nabla x^{a_r},
\eea
which is completely equivalent to the previous definition.  Applying the covariant curl to this expression gives us
\bea
\nabla \wedge \mv{\omega} &=& \frac{1}{r!} \left( \frac{\partial \omega_{a_1 a_2 \ldots a_r}}{\partial x^b} \right) \nabla x^b \wedge \nabla x^{a_1} \wedge \nabla x^{a_2} \wedge \ldots \wedge \nabla x^{a_r} \nonumber \\ &\qquad& + \ \frac{1}{r!} \omega_{a_1 a_2 \ldots a_r} \Big\{ \left( \nabla \wedge \nabla x^{a_1} \right) \wedge \nabla x^{a_2} \wedge \ldots \wedge \nabla x^{a_r} - \nabla x^{a_1} \wedge \left( \nabla \wedge \nabla x^{a_2} \right) \wedge \ldots \wedge \nabla x^{a_r} + \ldots \nonumber \\
 &\qquad& \qquad + \ (-1)^{r-1} \nabla x^{a_1} \wedge \nabla x^{a_2} \wedge \ldots \wedge \left( \nabla \wedge \nabla x^{a_r} \right) \Big\} ;
\eea
it is immediately evident that this is equivalent to the exterior derivative if and only if $\nabla \wedge \nabla x^a = 0$---which is true in the absence of torsion.

We can extend the development to arbitrary (noncoordinate) frames by transforming the arbitrary multivector with a vierbein (frame map) operator:
\bea
\adjinv{h}(\mv{\omega}) = \frac{1}{r!} \omega_{k_1 k_2 \ldots k_r} \hvec{e}^{k_1} \wedge \hvec{e}^{k_2} \wedge \ldots \wedge \hvec{e}^{k_r},
\eea
where
\bea
\hvec{e}^k = h^k_a \DD x^a = \adjinv{h}(\DD x^k ).
\eea
Applying the exterior derivative to this result, we get
\bea
\DD \adjinv{h}(\mv{\omega}) = \frac{1}{r!} \left( \frac{\partial}{\partial x^b} \omega_{k_1 k_2 \ldots k_r} h^{k_1}_{a_1} h^{k_2}_{a_2} \ldots h^{k_r}_{a_r} \right) \DD x^b \wedge \DD x^{a_1} \wedge \DD x^{a_2} \wedge \ldots \wedge \DD x^{a_r},
\eea
which follows directly from \eq{extdiff}.

\subsection{The Structure Equations without Torsion}

The orthonormal frame is defined in this way:
\bea
\hvec{e}^k \cdot \hvec{e}^l = \eta^{kl} = \adjinv{h}(\DD x^k) \cdot \adjinv{h}(\DD x^l) = h^k_\mu h^l_\nu g^{\mu\nu}.
\eea
Applying the covariant derivative to this expression yields (in terms of components)
\bea
\hvec{e}_i \cdot \nabla \hvec{e}^k \equiv - \omega^k_{ij} \hvec{e}^j,
\eea 
which, when protracted, gives us the connection 1-forms:
\bea
\nabla \wedge \hvec{e}^k = - \hvec{e}^i \wedge \left( \omega^k_{ij} \hvec{e}^j \right) \equiv - \vec{\omega}_j{}^k \wedge \hvec{e}^j.
\label{oneforms}
\eea
Putting all of this together gives the 1-forms in terms of the connection bivectors:
\bea
\vec{\omega}_j{}^k = \vec{\partial_a} \mv{\Omega}(\vec{a}) \cdot \left( \hvec{e}_j \wedge \hvec{e}^k \right).
\label{onetotwo}
\eea
Although it requires calculating the connection bivectors first, this expression is noteworthy as an explicit expression for the connection 1-forms.

Using the identification of covariant curl with exterior derivative, we get
\bea
\DD \hvec{e}^k = \nabla \wedge \hvec{e}^k - h^k_\mu \nabla \wedge \nabla x^\mu = - \vec{\omega}_j{}^k \wedge \hvec{e}^j;
\eea
which we immediately rearrange to yield Cartan's first structure equation:
\bea
\DD \hvec{e}^k + \vec{\omega}_j{}^k \wedge \hvec{e}^j = 0.
\label{cartan1}
\eea
Cartan's second structure equation is found in a similar way:
\bea
\DD \vec{\omega}_j{}^k + \vec{\omega}_l{}^k \wedge \vec{\omega}_j{}^l = \mv{\Theta}_j{}^k
\eea
where
\bea
\mv{\Theta}_j{}^k \equiv \tensor{R}(\hvec{e}_j \wedge \hvec{e}^k)
\eea
is the Riemann curvature tensor 2-form in the orthonormal frame.

From the preceeding, we see that differential geometry can be folded into the formalism of geometric algebra.  In addition, \eq{onetotwo} provides an expression relating the connection 1-forms of differential geometry (for which there is no direct expression for calculation that the authors are aware of) to the connection bivectors, which we can find explicitly using the techniques of Appendix~\ref{connection_derivation}.  It remains to be seen whether there is a useful expansion of \eq{onetotwo} in terms of differential forms.

\subsection{Example:  Connection 1-Forms for the Schwarzschild Case}

On the level of components, the connection 1-forms and bivectors are equivalent, so in a very real sense geometric algebra provides a more direct method to computing the connection than is provided by \eq{cartan1}.  In the absence of a more primitive form of \eq{onetotwo}, we use 
\beann
\vec{\omega}_j{}^k = \vec{\partial_a} \mv{\Omega}(\vec{a}) \cdot \left( \hvec{e}_j \wedge \hvec{e}^k \right) = \vec{g}^\mu \left\{ \mv{\Omega}_\mu \cdot \left( \hvec{e}_j \wedge \hvec{e}^k \right) \right\},
\eeann
where the $\mv{\Omega}_\mu$ are the connection bivectors in \eq{schwarzconnect}, to obtain
\bea
\begin{array}{lll}
\vec{\omega}_0{}^1 = \vec{\omega}_1{}^0 = g'_1 \hvec{e}_0 - g'_2 \hvec{e}_1 &\quad&
\vec{\omega}_0{}^2 = \vec{\omega}_2{}^0 = - g_2 \hvec{e}_2 /r \\
\vec{\omega}_0{}^3 = \vec{\omega}_3{}^0 = - g_2 \hvec{e}_3 /r &\quad&
\vec{\omega}_1{}^2 = - \vec{\omega}_2{}^1 = - g_1 \hvec{e}_2 /r \\
\vec{\omega}_1{}^3 = - \vec{\omega}_3{}^1 = - g_1 \hvec{e}_3 /r &\quad&
\vec{\omega}_2{}^3 = - \vec{\omega}_3{}^2 = - \cot \theta \hvec{e}_3 /r
\end{array}
\eea
These results reduce to the canonical expressions (see \emph{e.g.} \cite{wald}) used in obtaining the Schwarzschild solution when $g_2 = 0$, and agree with the expressions obtained using differential geometry directly \cite{zapolsky}.

\section{Null Vectors and Tetrads}
\label{null_tetrads}

Null vectors, especially null geodesics, often indicate symmetries in particular spacetime geometries.  In standard expositions \cite{exact,newman-penrose,janis-newman,penrose_rindler1}, a complex structure is often introduced to allow for four null vectors in a spacetime, and tensorial objects acquire complex character as well.  This is known as the \emph{null tetrad} formalism, and is intimately related to spinors.

Although we will not discuss spinors in this paper, null vectors and tetrads are useful to our general development.  An external complex structure is not needed in geometric algebra, since we have an object ready-made to fill the role of $i$:  the spacetime pseudoscalar.  We will ``complexify'' spacetime in the next section, then define null tetrads as mixed-grade multivectors in Section~\ref{nulltetradform}, which geometrically are Penrose's flags and flagpoles.  In Section~\ref{petrov_class}, we will find that the ``complex Weyl tensor'' is identical to the Weyl tensor we have already defined in Section~\ref{GR_in_GA}, and discuss the Petrov classification.  In this way, geometric algebra encompasses the complex structure often used in general relativity, elucidating the underlying geometric structures, without introducing any fundamentally new objects.

\subsection{Paravectors and ``Complex'' 4-Space}

As discussed in some detail in Appendix~\ref{even_subalgebra}, the space of scalars, pseudoscalars, and bivectors form an algebra under the geometric product.  This is called the \emph{even subalgebra} $\Cl^+_{1,3}$, and it is well-known that it is isomorphic to the three-dimensional geometric algebra $\Cl_3$.  We will find it useful in dealing with the Lorentz group \cite{lounesto}, and in constructing a $(3+1)$-dimensional ``complex'' space \cite{baylis,complex}.  The latter discussion will allow us to construct null tetrads, which in turn leads us to the Petrov classification of the Weyl tensor, which we will examine in Section~\ref{petrov_class}.  In a subsequent paper, we will examine the topic of 2-spinors in this space; see also \cite{jones} for a preliminary sketch of some of these ideas.

A useful way to obtain the even subalgebra requires selecting a particular timelike direction $\hvec{t}$ in the full algebra $\Cl_{1,3}$, where $\hvec{t}^2 = 1$ for simplicity \cite{STA}.  This allows us to define a \emph{paravector} $\pv{a}$ for any vector $\vec{a}$ in spacetime:
\bea
\vec{a} \ \to \ \pv{a} = \vec{a} \hvec{t} = a_0 +\mv{A}
\label{paravector}
\eea
where $a_0 = \vec{a}\cdot\hvec{t}$ and $\mv{A} = \vec{a}\wedge\hvec{t}$ define the scalar and timelike bivector parts of the paravector.  As discussed in Appendix~\ref{even_subalgebra}, the space formed from scalars, pseudoscalars, and bivectors is closed under the geometric product, so the product of paravectors with paravectors will always lie in this subalgebra.  

If we define the operation of \emph{Hermitian conjugation} for any multivector $\mv{Q}$ as
\bea
\mv{Q}^\dagger \equiv \hvec{t} \tilde{\mv{Q}} \hvec{t},
\eea
we see that the paravector we defined in \eq{paravector} is Hermitian:  $\pv{a}^\dagger = \pv{a}$.  In fact, the selection of the timelike direction allows us to split the even subalgebra into two pieces:  Hermitian, or ``real'', and anti-Hermitian, or ``imaginary''.  In terms of a generic multivector $\pv{\psi} \in \Cl^+_{1,3}$
\bea
\pv{\psi} = \grade{\pv{\psi}}_0 + \grade{\pv{\psi}}_2 + \grade{\pv{\psi}}_4
\eea
this split can be written as
\bea
\pv{\psi} = \grade{\pv{\psi}}_\Re + i \grade{\pv{\psi}}_\Im ,
\eea
where
\bea
\grade{\pv{\psi}}_\Re &\equiv& \frac{1}{2} \left( \pv{\psi} + \pv{\psi}^\dagger \right) \nonumber \\
\grade{\pv{\psi}}_\Im &\equiv& \frac{1}{2 i} \left( \pv{\psi} - \pv{\psi}^\dagger \right).
\eea
In the remainder of this paper, we will often talk about ``real'' and ``imaginary'' bivectors (meaning timelike and spacelike), and use the term ``complex number'' to mean the sum of a scalar and a pseudoscalar.  

For simplicity, let us define a (real) basis for $\Cl^+_{1,3}$ that contains the elements $\{1,\mv{e} = \vec{\sigma}_1,\mv{u}=\vec{\sigma}_2,\mv{v}=\vec{\sigma}_3\}$, where $\mv{e}^2 = \mv{u}^2 = \mv{v}^2 = 1$ are real bivectors.  We also require that the bivector basis be right-handed:  $\mv{e u} = -\mv{u e} = i \mv{v}$ and cyclic permutations, so that 
\bea
\vec{\sigma}_i \vec{\sigma}_j = \vec{\sigma}_i \cdot \vec{\sigma}_j + \vec{\sigma}_i \times \vec{\sigma}_j ,
\eea
which is to say that the outer product of real bivectors vanishes.  In this basis, a ``complex paravector'' is written
\bea
\pv{\psi} = \psi_0 + \psi_e \mv{e} + \psi_u \mv{u} + \psi_v \mv{v},
\label{diracbasis}
\eea
where the $\psi_a$ are ``complex'' scalars.  A complex paravector can therefore be seen as a four-dimensional complex vector, whose basis elements are the scalar and three real bivector basis elements, and whose components are the four complex (or eight real) scalars.  (Note that a complex paravector is isomorphic to a Dirac spinor in the language of Hestenes \cite{realspinors}; see also \cite{lounesto}, Chapter 10.)

Because paravectors are mixed-grade multivectors, the usual inner and outer products are often of limited usefulness; instead, we define (echoing Baylis \cite{baylis}) ``scalar'' and ``bivector'' products:
\begin{subequations}
\bea
\grade{\mv{AB}}_S = \grade{\mv{BA}}_S = \frac{1}{2} \left(\mv{A}\tilde{\mv{B}} + \mv{B}\tilde{\mv{A}} \right) \label{scalarprod} \\
\grade{\mv{AB}}_B = -\grade{\mv{BA}}_B = \frac{1}{2} \left(\mv{A}\tilde{\mv{B}} - \mv{B}\tilde{\mv{A}} \right)
\eea
\end{subequations}
where the result of \eq{scalarprod} is a complex number.  (Recall that $\tilde{\mv{A}}$ denotes the \emph{reverse} of a multivector.)  In the case of a ``real'' paravector, the scalar product returns the Minkowski metric:
\bea
\grade{\pv{a}\pv{b}}_S = a_0 b_0 - \mv{a}\cdot\mv{b} = \vec{a}\cdot\vec{b}.
\label{isometric}
\eea
Thus, paravectors can be used as a replacement for spacetime 4-vectors; for a formulation of relativistic electromagnetism entirely in the even subalgebra $\Cl^+_{1,3} \simeq \Cl_3$, see \cite{baylis}.

\subsection{Null Tetrads}
\label{nulltetradform}

In physical applications, null vectors often play an important role:  null geodesics identify paths for light, and can indicate symmetries in a system.  They are often used in obtaining the Petrov classification of the Weyl tensor, and are used in the Newman-Penrose method of finding exact solutions to Einstein's equations \cite{newman-penrose}, which we will not discuss in this paper.  A particularly useful way to work with null vectors is through complex null tetrads, which draws on the previous section.

If we let $\vec{N}^2 = 0$ be an arbitrary null vector, the paravector
\bea
\pv{N} = \vec{N} \hvec{t} = N_0 ( 1 + \mv{r} )
\eea
places all the relevant information in two quantities:  the ``extent'' $N_0 = \vec{N} \cdot \hvec{t}$, and the spatial direction $\mv{r}$, which is defined to be of unit length.  It is easily seen that $\grade{\pv{N}\pv{N}}_S = 0$, following from \eq{isometric}.  For the sake of discussion, we arbitrarily align $\mv{r}$ with $\mv{e}$ of the paravector basis from the previous section, and let
\bea
\pv{N} = 2 N_0 \pv{l}  \quad \text{where} \quad \pv{l} \equiv \frac{1}{2} ( 1 + \mv{e} ) .
\eea
The paravector $\pv{l}$ is called a \emph{flagpole}, which is intimately related to spinors (see \cite{jones} for more information).  It has a few other interesting properties, some of which we summarize here:
\bea
\begin{array}{lllllll}
\pv{l}^2 = \pv{l} &\quad& \text{idempotence} &\qquad&
\pv{l} + \tilde{\pv{l}} = 1 &\quad& \text{complementarity}\\
\pv{l}\tilde{\pv{l}} = 0 &\quad& \text{nullity} &\qquad&
\pv{l}^\dagger = \pv{l} &\quad& \text{Hermicity}
\end{array}
\eea
These properties indicate that $\pv{l}$ and $\tilde{\pv{l}}$ are linearly independent, so we define a new basis for paravectors around them:
\bea
\pv{l} &=& \frac{1}{2}(1+\mv{e}) \nonumber \\
\pv{n} &=& \tilde{\pv{l}} = \frac{1}{2}(1-\mv{e}) \nonumber \\
\pv{m} &=& \frac{1}{2}(1+\mv{e})\mv{u} = \frac{1}{2}(\mv{u} + i \mv{v}) \nonumber \\
\pv{m}^\dagger &=& \frac{1}{2}\mv{u}(1+\mv{e}) = \frac{1}{2}(\mv{u}-i \mv{v}).
\label{null_tetrad}
\eea
We call this set a \emph{null tetrad}, since its elements obey the following relation:
\bea
\pv{l}\tilde{\pv{l}} = \pv{n}\tilde{\pv{n}} = \pv{m}\tilde{\pv{m}} = \pv{m}^\dagger\tilde{\pv{m}}^\dagger = 0
\eea
and
\bea
\grade{\vec{ln}}_S = - \grade{\vec{mm}^\dagger}_S = \frac{1}{2} ,
\eea
with all other scalar products vanishing.  The first two tetrad elements in \eq{null_tetrad} are flagpoles, while the second two are their corresponding flags.  (Since we do not deal with spinors in this paper, the construction of tetrads is a trifle awkward.  See \cite{jones} for a better technique in geometric algebra, and \cite{penrose_rindler1} for the canonical discussion.)

We will occasionally find it useful to let $\pv{E}_0 \equiv \pv{l}$, $\pv{E}_1 \equiv \pv{n}$, $\pv{E}_2 \equiv \pv{m}$, and $\pv{E}_3 \equiv \pv{m}^\dagger$; in this notation, we see that the basis transforms as
\bea
\pv{E}'_a = \mv{R} \pv{E}_a \mv{R}^\dagger ,
\eea
where $\mv{R}$ is a rotor, as before.  We examine the transformation that preserves a particular null direction in Appendix~\ref{null_rot}.  The null tetrad ``metric'', which is invariant under Lorentz transformations, is
\bea
2 \grade{\pv{E}_a \pv{E}_b}_S \equiv \eta_{ab}
\eea
where \cite{newman-penrose}
\bea
\left[ \eta_{ab} \right] = \left( \begin{array}{cccc} 0 & 1 & 0 & 0 \\ 1 & 0 & 0 & 0 \\ 0 & 0 & 0 & -1 \\ 0 & 0 & -1 & 0 \end{array} \right) .
\eea
A general multivector is then written as
\bea
\vec{A} = A^a \pv{E}_a,
\eea
where the $A^a$ are complex scalars---demonstrating that the null tetrads form a basis for $\Cl^+_{1,3}$ over the complex numbers.

The bivector basis of \cite{exact} (with different indexing, however) can be reproduced by taking the nonzero vector products of the tetrads:
\bea
\mv{Z}_0 &\equiv& - \sqrt{2} \grade{\pv{l}\pv{m}}_B = \frac{1}{\sqrt{2}} (1+\mv{e}) \mv{u} \nonumber \\
\mv{Z}_1 &\equiv& \grade{\pv{l}\pv{n}}_B - \grade{\pv{m}\pv{m}^\dagger}_B = \mv{e} \nonumber \\
\mv{Z}_2 &\equiv& - \sqrt{2} \grade{\pv{n}\pv{m}^\dagger}_B = \frac{1}{\sqrt{2}}(1-\mv{e})\mv{u}
\label{bivbasis}
\eea
These are linearly-independent, and span a three-dimensional space over the complexes.  We can write a ``metric'' for them:
\bea
\grade{\mv{Z}_a\mv{Z}_b}_S = z_{ab} \quad \to \quad \left[ z_{ab} \right] = \left( \begin{array}{ccc} 0 & 0 & -1 \\ 0 & -1 & 0 \\ -1 & 0 & 0 \end{array} \right)
\label{bivmetric}
\eea
which is obviously real-valued, despite the ``complex'' character of the $\mv{Z}_a$ themselves.  Thus, when we write a bivector
\bea
\vec{A} = A^a \mv{Z}_a
\eea
the complex components $A^a$ are easily extracted using the scalar product.  And, unlike in the standard formalism, the bivector basis is easily interpreted in geometric algebra:  $\mv{Z}_0$ and $\mv{Z}_2$ are oppositely-directed null flags determined by the timelike direction $\mv{Z}_1 = \mv{e}$.  We will see further examples of the new geometric interpretations that geometric algebra provides in the next section.

\section{Classification of the Weyl Tensor}
\label{petrov_class}

This section is concerned not with finding solutions to the field equations, but rather extracting geometrical information about the Weyl conformal tensor.  In the case of vacuum solutions, the Weyl tensor is the gravitational field, so being able to discuss the form of the solution \emph{without} first solving the equations can help us understand the underlying physical configuration.  The classification of the Weyl tensor due to Petrov, which has been translated into the language of geometric algebra by Sobczyk \cite{STAC,complex,classification} and Hestenes and Sobczyk \cite{CA2GC}, provides one such way of describing the fields, independent of a coordinate system.  We examine an algebraic technique using the eigenvalue problem in Section~\ref{petrov_eigen}, then we apply the null tetrad formalism in Section~\ref{geomclass} to provide a geometric interpretation of certain of these classes.  This last discussion will result in illuminating new canonical forms for the gravitational field.

The Weyl tensor is symmetric
\bea
\mv{A}\cdot\tensor{C}(\mv{B}) = \tensor{C}(\mv{A})\cdot\mv{B}
\label{weylsymm}
\eea
and self-dual
\bea
\tensor{C}(i\mv{B}) = i \tensor{C}(\mv{B}),
\label{weyldual}
\eea
where $\mv{A}$ and $\mv{B}$ are bivectors.  Combining these properties yields the important result \cite{STAC}
\bea
\grade{\mv{A}\tensor{C}(\mv{B})}_S = \grade{\mv{B}\tensor{C}(\mv{A})}_S.
\label{dualsymm}
\eea
We draw on an electromagnetic analogue by writing \cite{exact}
\bea
\tensor{C}(\mv{B}) = \tensor{E}(\mv{B}) + i \tensor{B}(\mv{B})
\eea
where $\tensor{E}(\mv{B})$ and $\tensor{B}(\mv{B})$ are both real bivector-valued symmetric operators.  Because of the self-duality of the full Weyl tensor, we see that
\bea
\tensor{E}(i\mv{B}) = - \tensor{B}(\mv{B}) \quad \text{and} \quad \tensor{B}(i\mv{B}) = \tensor{E}(\mv{B}).
\eea
These results will be used extensively in the following sections.

\subsection{The Petrov Classification:  The Eigenvalue Problem}
\label{petrov_eigen}

The Petrov type is most easily determined by solving the eigenvalue problem:
\bea
\tensor{C} (\mv{B}) - \lambda \mv{B} = 0
\eea
where $\lambda$ is a complex scalar.  The Weyl tensor can be viewed as a $3\times 3$ complex symmetric matrix acting on the 3-dimensional complex space of bivectors.  Since the operator $\tensor{C} - \lambda \tensor{1}$ is zero, it is singular, and we are left with the usual determinant
\bea
\left| \tensor{C} - \lambda \tensor{1} \right| = 0 .
\eea
If we define a timelike (real) bivector basis $\{\vec{\sigma}_k\}$ (for $k = 1,2,3$), we can write the determinant as \cite{STAC}
\bea
\left| \tensor{C} - \lambda \tensor{1} \right| = - \vec{\sigma}_1\vec{\sigma}_2\vec{\sigma}_3 \grade{ \left[ \left( \tensor{C}(\vec{\sigma}_1) - \lambda \vec{\sigma}_1 \right) \times \left( \tensor{C}(\vec{\sigma}_2) - \lambda \vec{\sigma}_2 \right) \right] \left( \tensor{C}(\vec{\sigma}_3) - \lambda \vec{\sigma}_3 \right) }_S = 0
\label{characteristic}
\eea
which is a third-order algebraic equation in $\lambda$.  Although this expression looks hideously complicated, we will shortly calculate it for the Schwarzschild case, and show that it is actually remarkably simple to work out.  (See also \cite{classical} for a worked example in $\Cl_3$.)  Because the Weyl tensor is tractionless
\bea
\vec{\partial_a} \tensor{C}(\vec{a}\wedge\vec{b}) = 0 \quad \forall \vec{b}
\eea
the eigenvalues satisfy the condition
\bea
\lambda_1 + \lambda_2 + \lambda_3 = 0,
\label{eigennull}
\eea
which means only two eigenvalues are independent.

Stated simply, the eigenbivectors determine the general Petrov type (labeled I, II, and III), while specific eigenvalue restrictions yield special cases (types D and N).  %We take the eigenvalue equation, and expand each eigenbivector $\mv{B}_a$ into its components with respect to the basis:
%\bea
%\tensor{C} (\mv{B}_a) - \lambda_a \mv{B}_a = B^k_a \left( \tensor{C}(\vec{\sigma}_k) - \lambda_a \vec{\sigma}_k \right) \equiv B^k_a \vec{c}_{a, k} = 0
%\eea
%where the $B^k_a$ are complex scalars, while the index $a$ indicates the choice of eigenvalue and is not summed.  Taken as an algebraic equation in the components $B^k_a$, this is severely underdetermined, so we eliminate one component arbitrarily by applying the commutator product:
%\bea
%(B^k_a \vec{c}_{a, k}) \times \vec{c}_{a, 3} = B^1_a \vec{c}_{a,1} \times \vec{c}_{a,3} + B^2_a \vec{c}_{a,2} \times \vec{c}_{a,3} = 0 .
%\eea
%This allows us to find ratios of components, and hence the eigenbivectors.  
Petrov type I occurs when the eigenbivectors span the entire 3-dimensional space; the bivectors for types II and III span a two- and one-dimensional subspace, respectively, and both have a null eigvenbivector:
\beann
\mv{B}^2_1 = 0 \qquad \text{(types II, N, and III)}
\eeann
The last two types are special cases:  D corresponds to type I when two eigenvalues are equal, while type N corresponds to type II with all three eigenvalues vanishing.  Geometric interpretations are left to Section~\ref{geomclass}.

Hestenes and Sobczyk \cite{CA2GC} have shown that the Weyl tensor can be written as
\bea
\tensor{C} (\mv{X}) = Q^{jk} \vec{\sigma}_j \mv{X} \vec{\sigma}_k
\label{HSweyl}
\eea
where the symmetric matrix $Q^{jk}=Q^{kj}$ is comprised of complex scalars.  However, with the eigenbivectors in hand for types I and II (which include types D and N), we can rewrite this as
\bea
\tensor{C} (\mv{X}) = \sum^2_{k=1} \left\{ \mv{B}_k \mv{X} \mv{B}_k + \frac{1}{3} \mv{X} \mv{B}^2_k \right\} \mu_k \qquad \text{(types I, II, D, and N)}
\label{canonical}
\eea
where $\mu_1 = \lambda_1 + \lambda_2 /2$ and $\mu_2 = \lambda_2 + \lambda_1 /2$ are linear combinations of the eigenvalues.  (For type I, the third eigenvalue is found from \eq{eigennull}, and the third eigenvector is not needed for the calculation.)  Type D corresponds either to $\mu_1 = \mu_2$ or $\mu_2 = 0$, while type N corresponds to $\mu_2 = 0$ when $\mv{B}_1$ is the null eigenbivector.  Type III has the canonical form
\bea
\tensor{C} (\mv{X}) = \left( \mv{B} \mv{X} \mv{F} + \mv{F} \mv{X} \mv{B} \right) \mu \qquad \text{(type III)}
\label{canonical3}
\eea
where $\mu$ is a complex scalar, and the bivector $\mv{F}$ is orthogonal to the eigenbivector:  $\mv{F}\cdot \mv{B} = 0$.

\subsection{Geometric Classification Using Tetrad Techniques}
\label{geomclass}

The forms for the Weyl tensor obtained in the previous section are useful, but we can go a step further.  Motivated by the Newman-Penrose formulation, we use the tetrads and the bivector basis constructed from them to understand what the types mean geometrically.  Although Szekeres obtained the same qualitative results \cite{compass}, here we provide new canonical forms for the Weyl tensor.

We begin by using \eq{dualsymm} to define the complex scalar-valued function
\bea
\Psi(\mv{A},\mv{B}) \equiv - \grade{\mv{A} \tensor{C}(\mv{B})}_S = \mv{A}\cdot\tensor{C}(\mv{B}) + \mv{A}\wedge\tensor{C}(\mv{B}).
\eea
When we use the bivector basis defined in \eq{bivbasis}, we get 
\bea
\Psi_{ab} = - \grade{\mv{Z}_a \tensor{C}(\mv{Z}_b)}_S = \Psi_{ba},
\eea
which can be regarded as the components of a $3\times 3$ symmetric, complex matrix.  We reduce the amount of redundancy by labeling the components \cite{newman-penrose}
\bea
\Psi_0 &\equiv& \Psi_{00} = -\grade{\mv{Z}_0 \tensor{C}(\mv{Z}_0)}_S \nonumber \\
\Psi_1 &\equiv& \Psi_{01} = \Psi_{10} = -\grade{\mv{Z}_0 \tensor{C}(\mv{Z}_1)}_S \nonumber \\
\Psi_2 &\equiv& \Psi_{11} - \Psi_{02} = \Psi_{11} - \Psi_{20} = - \grade{\mv{Z}_1 \tensor{C}(\mv{Z}_1)}_S + \grade{\mv{Z}_0 \tensor{C}(\mv{Z}_2)}_S \nonumber \\
\Psi_3 &\equiv& \Psi_{12} = \Psi_{21} = - \grade{\mv{Z}_1 \tensor{C}(\mv{Z}_2)}_S \nonumber \\
\Psi_4 &\equiv& \Psi_{22} = -\grade{\mv{Z}_2 \tensor{C}(\mv{Z}_2)}_S .
\label{Psi_matrix}
\eea
The sixth component, which has been folded into $\Psi_2$,
\bea
\Psi_{02} = \Psi_{20} = -\grade{\mv{Z}_0 \tensor{C}(\mv{Z}_2)}_S
\eea
is generally not talked about in algebraic discussions of the Weyl tensor; this is because it is not an independent quantity, due to the trace-free property \cite{exact}.  In certain cases, it is useful to fix one null direction and rotate the others, thereby obtaining new values for some of the $\Psi_a$; Appendix~\ref{null_rot} is devoted to that topic.

To perform certain calculations, we will find it useful to write the complex bivector basis $\{\mv{Z}_a\}$ in terms of the real bivector basis $\{\mv{e},\mv{u},\mv{v}\}$:
\bea
\mv{e} = \mv{Z}_1 \quad \mv{u} = \frac{1}{\sqrt{2}} (\mv{Z}_0 + \mv{Z}_2) \quad \mv{v} = \frac{1}{i \sqrt{2}} (\mv{Z}_0 - \mv{Z}_2),
\eea
and we split the tensor into its electric and magnetic parts and consider the quantities
\begin{subequations}
\bea
\Psi_{0b} &=& -\grade{\mv{Z}_0 \tensor{C}(\mv{Z}_b)}_S = 2^{-1/2} \left( E_u (\mv{Z}_b) - B_v (\mv{Z}_b) + i \left\{ E_v(\mv{Z}_b) + B_u(\mv{Z}_b)\right\}\right) \\
\Psi_{1b} &=& -\grade{\mv{Z}_1 \tensor{C}(\mv{Z}_b)}_S = E_e (\mv{Z}_b) + i B_e (\mv{Z}_b) \\
\Psi_{2b} &=&  -\grade{\mv{Z}_2 \tensor{C}(\mv{Z}_b)}_S = 2^{-1/2} \left( E_u (\mv{Z}_b) + B_v (\mv{Z}_b) - i \left\{ E_v(\mv{Z}_b) - B_u(\mv{Z}_b)\right\}\right)
\eea
\label{reduction_psi}
\end{subequations}
along with the symmetry conditions
\begin{subequations}
\bea
\label{psi01}
\Psi_{01} = \Psi_{10} &\longrightarrow& \left\{ \begin{array}{l} \sqrt{2} E_e (\mv{Z}_0) = E_u (\mv{Z}_1) - B_v (\mv{Z}_1) \\ 
\sqrt{2} B_e (\mv{Z}_0) = E_v (\mv{Z}_1) + B_u (\mv{Z}_1) \end{array} \right. \\ 
\label{psi02}
\Psi_{02} = \Psi_{20} &\longrightarrow& \left\{ \begin{array}{l} E_u (\mv{Z}_2) - B_v(\mv{Z}_2) = E_u (\mv{Z}_0) + B_v(\mv{Z}_0) \\ 
E_v (\mv{Z}_2) + B_u(\mv{Z}_2) = - E_v (\mv{Z}_0) + B_u(\mv{Z}_0) \end{array} \right. \\
\Psi_{12} = \Psi_{21} &\longrightarrow& \left\{ \begin{array}{l} \sqrt{2} E_e (\mv{Z}_2) = E_u (\mv{Z}_1) + B_v (\mv{Z}_1) \\ 
\sqrt{2} B_e (\mv{Z}_2) = - E_v (\mv{Z}_1) + B_u (\mv{Z}_1) \end{array} \right.
\eea
\label{symm_conditions}
\end{subequations}
In the preceeding, $E_e (\mv{Z}_b) \equiv \mv{e}\cdot \tensor{E}(\mv{Z}_b)$, and so forth.

We will use these decompositions to obtain the Petrov types for the various $\Psi_a$ in the following subsections.

\subsubsection{Transverse Waves:  $\Psi_0 \neq 0$ or $\Psi_4 \neq 0$}

In the case where $\Psi_0 \neq 0$ but the rest vanish, we immediately get (from \eq{reduction_psi})
\beann
E_e (\mv{Z}_b) = B_e (\mv{Z}_b) = 0,
\eeann
while applying \eq{symm_conditions} yields
\beann
\tensor{C}(\mv{Z}_1) = \tensor{C}(\mv{Z}_2)= 0
\eeann
and
\beann
B_v(\mv{Z}_0) &=& - E_u(\mv{Z}_0) \nonumber \\
E_v(\mv{Z}_0) &=& B_u (\mv{Z}_0),
\eeann
so the electric and magnetic pieces are perpendicular to each other.  Putting all of this together, we see that
\beann
\Psi_0 &=& \sqrt{2} \left( E_u (\mv{Z}_0) - i B_u(\mv{Z}_0) \right)
\eeann
and
\beann
\tensor{E}(\mv{Z}_0) \cdot \tensor{B}(\mv{Z}_0) = E_u (\mv{Z}_0) B_u (\mv{Z}_0) + E_v (\mv{Z}_0) B_v (\mv{Z}_0) = E_u (\mv{Z}_0) B_u (\mv{Z}_0) - B_u(\mv{Z}_0) E_u (\mv{Z}_0) = 0.
\eeann
We can then write the nonvanishing piece of the Weyl tensor as
\beann
\tensor{C}(\mv{Z}_0) = E_u \mv{Z}_2 = - B_v \mv{Z}_2 , \qquad \Psi_0 \neq 0 ,
\eeann
or, utilizing \eq{bivmetric}, for an arbitrary bivector $\mv{X} = X^a \mv{Z}_a$
\bea
\tensor{C}(\mv{X}) = - E_u \mv{Z}_2 \grade{ \mv{Z}_2 \mv{X} }_S = \frac{E_u}{2} \mv{Z}_2 \mv{X} \mv{Z}_2 .
\label{psi0}
\eea
Similarly, when $\Psi_4 \neq 0$ but the rest vanishing we get
\bea
\Psi_{a\neq 4} = 0 \ \to \ \tensor{C}(\mv{X}) = \frac{E_u}{2} \mv{Z}_0 \mv{X} \mv{Z}_0 , \quad \tensor{C}(\mv{Z}_0) = \tensor{C}(\mv{Z}_1) = 0 .
\label{psi4}
\eea
Equations (\ref{psi0}) and (\ref{psi4}) already are in the canonical form for Petrov type N; confirming this classification is a simple matter, using \eq{characteristic} in the form
\bea 
\left| \tensor{C} - \lambda \tensor{1} \right| = - \mv{e}\mv{u}\mv{v} \grade{ \left[ \left( \tensor{C}(\mv{e}) - \lambda \mv{e} \right) \times \left( \tensor{C}(\mv{u}) - \lambda \mv{u} \right) \right] \left( \tensor{C}(\mv{v}) - \lambda \mv{v} \right) }_S = 0 ,
\label{chartrans}
\eea
and using the canonical forms to evaluate the expression.

The Weyl tensor represented in both cases here is proportional to a null bivector.  Its electric and magnetic pieces are perpendicular to each other, and both are perpendicular to the direction $\mp\mv{e}$.  This means that the Weyl tensor represents a transverse plane wave traveling in the $(\mp\mv{e})$-direction, which agrees with the analysis by Szekeres \cite{compass} (albeit by a different route); see \cite{baylis} for the electromagnetic analogue.

\subsubsection{Coulombic Case:  $\Psi_2 \neq 0$}

The final case which is directly related to the electromagnetic treatment is when $\Psi_{a \neq 2} = 0$.  Here, we find immediately that
\bea
\tensor{C}(\mv{Z}_1) = \left(E_e(\mv{Z}_1) + i B_e (\mv{Z}_1)\right) \mv{Z}_1 \equiv F_1 \mv{Z}_1,
\eea
while
\beann
\tensor{C}(\mv{Z}_0) &=& \left( E_u(\mv{Z}_0) + i B_u(\mv{Z}_0)\right) \mv{Z}_0 \equiv F_2 \mv{Z}_0 \nonumber \\
\tensor{C}(\mv{Z}_2) &=& \left( E_u(\mv{Z}_0) + i B_u(\mv{Z}_0)\right) \mv{Z}_2 \equiv F_2 \mv{Z}_2 ,
\eeann
where we have exploited the symmetry conditions to get
\beann
E_e (\mv{Z}_0) = E_e (\mv{Z}_2) = B_e(\mv{Z}_0) = B_e (\mv{Z}_2) &=& 0 \nonumber \\
E_u (\mv{Z}_1) = E_v (\mv{Z}_1) = B_u (\mv{Z}_1) = B_v (\mv{Z}_1) &=& 0
\eeann
and
\beann
E_u(\mv{Z}_2) = E_u(\mv{Z}_0) \quad B_u(\mv{Z}_2) = B_u(\mv{Z}_0).
\eeann

The complex scalars $F_a$ are obviously the eigenvalues for the system, which we can confirm by putting all of this into \eq{chartrans}:
\beann
\lambda_1 = E_e (\mv{Z}_1) + i B_e (\mv{Z}_1) = F_1 , \quad \lambda_2 = \lambda_3 = E_u(\mv{Z}_0) + i B_u(\mv{Z}_0) = F_2 .
\eeann
Enforcing \eq{eigennull} further restricts the eigenvalues:
\beann
\sum^3_{k=1} \lambda_k = 0 \quad \to \quad E_u (\mv{Z}_0) = - \frac{1}{2} E_e(\mv{Z}_1) \quad \text{and} \quad B_u (\mv{Z}_0) = - \frac{1}{2} B_e(\mv{Z}_1).
\eeann
This is Petrov type D, with eigenvectors $\mv{Z}_b$ (or equivalently $\{\mv{e},\mv{u},\mv{v}\}$).  This means that all incoming information is balanced by outgoing, if one regards $\mv{e}$ as a ``radial'' direction---hence, type D tensors are ``Coulombic'' \cite{janis-newman}.

We obtain the canonical form by noting that
\beann
\tensor{C}(\mv{X}) = - F_1 \mv{Z}_1 \grade{ \mv{X} \mv{Z}_1 }_S - F_2 \Big( \mv{Z}_0 \grade{ \mv{X} \mv{Z}_2 }_S + \mv{Z}_2 \grade{ \mv{X} \mv{Z}_0 }_S \Big)  = \frac{1}{2} F_1 \left\{ \mv{Z}_1 \mv{X} \mv{Z}_1 - \frac{1}{2} \left( \mv{Z}_0 \mv{X} \mv{Z}_2 + \mv{Z}_2 \mv{X} \mv{Z}_0 \right) \right\} .
\eeann
Since $\mv{Z}^2_0 = \mv{Z}^2_2 = 0$ and $\mv{Z}_1$ anticommutes with both $\mv{Z}_0$ and $\mv{Z}_2$, we discover the identity
\beann
\frac{1}{2} \left( \mv{Z}_0 \mv{X} \mv{Z}_2 + \mv{Z}_2 \mv{X} \mv{Z}_0 \right) = - X^a \frac{1}{2} \delta_{a 1} \mv{Z}_1 \left( \mv{Z}_0 \mv{Z}_2 + \mv{Z}_2 \mv{Z}_0 \right) = \grade{ \mv{X} \mv{Z}_1 }_S \mv{Z}_1 ,
\eeann
so that
\bea
\tensor{C}(\mv{X}) = \frac{1}{2} F_1 \left\{ \mv{Z}_1 \mv{X} \mv{Z}_1 - \grade{ \mv{X} \mv{Z}_1 }_S \mv{Z}_1 \right\} = \frac{3}{4} F_1 \left\{ \mv{Z}_1 \mv{X} \mv{Z}_1 + \frac{1}{3} \mv{X} \right\},
\label{canonical_D}
\eea
which is the canonical form for Petrov type D.

\subsubsection{Longitudinal Waves:  $\Psi_1 \neq 0$ or $\Psi_3 \neq 0$}

For $\Psi_1 \neq 0$, we find that
\beann
\Psi_1 &=& E_e (\mv{Z}_0) + i B_e (\mv{Z}_0)
\eeann
using the same techniques as before, while the rest are zero; the Weyl tensor is then
\bea
\tensor{C}(\mv{Z}_0) &=& (E_e(\mv{Z}_0) + i B_e(\mv{Z}_0) ) \mv{Z}_1 \equiv F \mv{Z}_1 \nonumber \\
\tensor{C}(\mv{Z}_1) &=& (E_e(\mv{Z}_0) + i B_e(\mv{Z}_0) ) \mv{Z}_2 \equiv F \mv{Z}_2 \nonumber \\
\tensor{C}(\mv{Z}_2) &=& 0 .
\eea
$\tensor{C}(\mv{Z}_0)$ is solely in the $\mv{Z}_1$-direction, while $\tensor{C}(\mv{Z}_1)$ is a transverse piece---so this object has a longitudinal component \cite{compass}.  The characteristic equation gives us three zero eigenvalues, and since $\mv{Z}_2$ is the only eigenbivector, we have Petrov type III.  The canonical form is found immediately to agree with that analysis:
\bea
\tensor{C}(\mv{X}) = - F \left( \grade{ \mv{X} \mv{Z}_2}_S \mv{Z}_1 + \grade{ \mv{X} \mv{Z}_1 }_S \mv{Z}_2 \right) = \frac{F}{2} \left( \mv{Z}_2 \mv{X} \mv{Z}_1 + \mv{Z}_1 \mv{X} \mv{Z}_2 \right).
\eea
Repeating the process for $\Psi_3 \neq 0$ is trivial.

\subsection{Example:  The Petrov Type for the Schwarzschild Case}

For simplicity, we will use the orthonormal frame from Section~\ref{schwarz_geom} to define a natural bivector basis for the Schwarzschild solution:
\beann
\vec{\sigma}_k = \hvec{e}_k \hvec{e}_0 ,
\eeann
where $k$ runs from 1 to 3, and $\vec{\sigma}_1\vec{\sigma}_2\vec{\sigma}_3 = i$.  In this frame, the curvature bivectors are
\beann
\tensor{R}(\vec{\sigma}_1) = \tensor{C}(\vec{\sigma}_1) = -\frac{1}{2} \chi''(r) \vec{\sigma}_1 \ , \quad 
\tensor{R}(\vec{\sigma}_2) = \tensor{C}(\vec{\sigma}_2) = \frac{1}{2 r} \chi'(r) \vec{\sigma}_2 \ , \quad 
\tensor{R}(\vec{\sigma}_3) = \tensor{C}(\vec{\sigma}_3) = \frac{1}{2 r} \chi'(r) \vec{\sigma}_3
\eeann
where we have used the vacuum equations to set the Riemann tensor equal to the Weyl tensor; recall that $\chi(r) = 1 - 2 M/r$.  The eigenvalues and eigenvectors of $\tensor{C}$ can be read off immediately; however, for the sake of example we will evaluate \eq{characteristic} to obtain them.

The characteristic equation requires evaluation of the quantity
\beann
\left( \tensor{C}(\vec{\sigma}_1) - \lambda \vec{\sigma}_1 \right) \times \left( \tensor{C}(\vec{\sigma}_2) - \lambda \vec{\sigma}_2 \right) &=& \left( -\frac{1}{2} \chi''(r) - \lambda \right) \left( \frac{1}{2 r} \chi'(r) - \lambda \right) \vec{\sigma}_1 \times \vec{\sigma}_2 ;
\eeann
noting that $\vec{\sigma}_1 \times \vec{\sigma}_2 = i \vec{\sigma}_3$, we get
\beann
\left| \tensor{C} - \lambda \tensor{1} \right| &=& -i \left( -\frac{1}{2} \chi''(r) - \lambda \right) \left( \frac{1}{2 r} \chi'(r) - \lambda \right)^2 \grade{ i \vec{\sigma}_3 \vec{\sigma}_3 }_S 
= \left( -\frac{1}{2} \chi''(r) - \lambda \right) \left( \frac{1}{2 r} \chi'(r) - \lambda \right)^2 =0 .
\eeann
Solving this gives the eigenvalues
\beann
\lambda_1 = -\frac{1}{2} \chi''(r) = - \frac{2 M}{r^3}, \ \lambda_2 = \lambda_3 = \frac{1}{2 r} \chi'(r) = \frac{M}{r^3};
\eeann
two are repeated, so this is automatically either type D or (if $\chi$ is constant) type 0, which is the trivial Minkowski case.  The eigenbivectors are
\beann
\mv{B}_k = \vec{\sigma}_k,
\eeann
which spans the full three-dimensional complex space.

From \eq{canonical_D}, we see that the canonical form of the Weyl tensor is \cite{GTG}
\bea
\tensor{C} (\mv{X}) = - \frac{3 M}{2 r^3} \left( \vec{\sigma}_1 \mv{X} \vec{\sigma}_1 + \frac{1}{3} \mv{X} \right)
\eea
since $\mu_1 = \lambda_1 + 1/2 \lambda_2 = - (3 M)/(2 r^3)$ and $\mu_2 = \lambda_2 + 1/2 \lambda_1 = 0$, and $\sigma^2_k = 1$.  The eigenvalues are real, so there is no ``magnetic'' piece:  the nonrotating black hole solution is purely ``electric''.  As we will see in Section \ref{kerrsol}, the Kerr solution, which describes a rotating black hole, does contain a magnetic part, and in fact a trick exists to obtain it from the Schwarzschild solution by replacing the radial coordinate $r$ with $r - i a \cos \theta$ \cite{GTG,ks2}.

\section{Example:  The Kerr Solution}
\label{kerrsol}

The Schwarzschild metric is among the simplest solutions to Einstein's equations, but physical black holes will generally have nonvanishing angular momentum.  A more realistic---and more complicated---solution is the Kerr metric.  No truly simple techniques exist to obtain this metric, but a straightfoward method has been developed by Schiffer \emph{et al.} \cite{schiffer}, and adapted to the language of geometric algebra by Doran and colleagues \cite{ks1,ks2}, along with Snygg \cite{snygg}.  We will follow their path, showing the power of geometric algebra by performing coordinate-independent calculations, while providing some physical insight into the meaning of the ansatz used.

\subsection{The Kerr-Schild Ansatz and Curvature}

The Kerr-Schild ansatz, which is used by \cite{snygg,ks1,ks2,chandra,schiffer}, is defined in vierbein language as
\bea
\inv{h}(\vec{a}) = \vec{a} + M V \vec{a}\cdot \vec{l} \vec{l},
\eea
where $\vec{l}^2 = 0$ is an unspecified null vector, $M$ is identified with the black hole mass, and $V$ is a scalar function.  From this ansatz, we can define a new frame and its reciprocal from an arbitrary orthonormal frame
\bea
\vec{g}_\mu &=& \inv{h}(\hvec{e}_\mu) = \hvec{e}_\mu + M V \hvec{e}_\mu \cdot \vec{l} \vec{l} \nonumber \\
\vec{g}^\mu &=& \adj{h}(\hvec{e}^\mu) = \hvec{e}^\mu - M V \hvec{e}^\mu \cdot \vec{l} \vec{l}.
\label{ksansatz}
\eea
From \eq{ksansatz}, we obtain the usual Kerr-Schild metric
\bea
g_{\mu\nu} = \vec{g}_\mu \cdot\vec{g}_\nu = \eta_{\mu\nu} + 2 M V \hvec{e}_\mu \cdot \vec{l} \hvec{e}_\nu \cdot \vec{l}.
\label{ksmetric}
\eea
Two other features of the KS vierbein will prove to be important for us:
\bea
\adj{h}(\vec{l}) = \inv{h}(\vec{l}) = \vec{l} \quad \text{and} \quad \inv{h}(\vec{a}) \vec{l} = \adj{h}(\vec{a}) \vec{l} = \vec{a} \vec{l},
\eea
which means that $\vec{l}$ has the same components in both frames---
\beann
\vec{l} = \vec{g}^\mu \vec{g}_\mu \cdot \vec{l} = \hvec{e}^\mu \hvec{e}_\mu \cdot \vec{l}
\eeann
---while
\bea
0 = \vec{a} \cdot \hat{\nabla} (\vec{l}\cdot\vec{l}) = 2 \vec{l} \cdot \left( \vec{a} \cdot \hat{\nabla} \vec{l} \right) .
\label{nullity}
\eea

This ansatz is also useful for simplifying the calculation of connection bivectors:  using \eq{final_expression} in Appendix~\ref{connection_derivation}, we see that the third term vanishes, while the first two terms can be rewritten as
\bea
\mv{\Omega}(\vec{b}) = - M \left( \hat{\nabla} \wedge (V \vec{l} \cdot \vec{b} \vec{l} ) - V \vec{g}^\lambda \wedge \vec{l} \vec{l} \cdot \left( \hat{\nabla}_\lambda \vec{b} \right)\right) .
\eea
From this and \eq{nullity}, we see immediately that $\mv{\Omega}(\vec{l}) = 0$, while if $\vec{b} = \vec{g}_\alpha$ or $\vec{b} = \hvec{e}_\alpha$, the second term vanishes:
\bea
\mv{\Omega}_\mu = \mv{\Omega}(\vec{g}_\mu ) = \mv{\Omega}(\hvec{e}_\mu ) = - M \hat{\nabla} \wedge (V l_\mu \vec{l} ),
\eea
using the shorthand $l_\mu = \hvec{e}_\mu \cdot \vec{l} = \vec{g}_\mu \cdot \vec{l}$.  If we wish to write everything out in terms of the orthonormal frame, we get another interesting expression:
\bea
\mv{\Omega}(\vec{b}) = M \left\{ M V^2 \vec{l} \cdot \vec{b} \vec{l} \wedge \vec{v} + V \hvec{e}^\lambda \wedge \vec{l} \vec{l} \cdot \left( \hat{\nabla}_\lambda \vec{b} \right) - \hvec{e}^\lambda \wedge \hat{\nabla}_\lambda \left( V \vec{l} \cdot \vec{b} \vec{l} \right) \right\}
\eea
where
\bea
\vec{v} \equiv \vec{l} \cdot \nabla \vec{l} = \vec{l} \cdot \hat{\nabla} \vec{l}
\eea
is a vector that vanishes when $\vec{l}$ is geodesic; note that $\vec{l} \cdot \vec{v} = 0$ via \eq{nullity}.

Due to the simple form of the connection bivector, we can write down the curvature tensor involving $\vec{l}$ immediately:
\bea
\tensor{R}(\vec{g}_\mu \wedge \vec{l}) = - \vec{l} \cdot \hat{\nabla} \mv{\Omega}_\mu %\quad \to \quad \tensor{R}(\vec{a} \wedge \vec{l}) = - \vec{l} \cdot \hat{\nabla} \mv{\Omega}(\vec{a}) + \mv{\Omega}\left( \vec{l} \cdot \hat{\nabla} \vec{a} \right).
\eea
Following \cite{ks1} and \cite{chandra}, we form the quantity
\bea
\vec{l} \cdot \tensor{R} (\vec{l}) = \vec{l} \cdot \left( \vec{g}^\mu \cdot \tensor{R}(\vec{g}_\mu \wedge \vec{l}) \right)
\eea
where
\bea
\vec{g}^\mu \cdot \tensor{R}(\vec{g}_\mu \wedge \vec{l}) &=& - \vec{g}^\mu \cdot \left\{ \vec{l} \cdot \hat{\nabla} \mv{\Omega}_\mu \right\} = - \vec{l} \cdot \hat{\nabla} \left( \vec{g}^\mu \cdot \mv{\Omega}_\mu \right) + \left( \vec{l} \cdot \hat{\nabla} \vec{g}^\mu \right) \cdot \mv{\Omega}_\mu
\eea
is the Ricci tensor $\tensor{R}(\vec{l})$.  Noting that
\begin{subequations}
\bea
- M^{-1} \vec{g}^\mu \cdot \mv{\Omega}_\mu &=& V \vec{v} + \vec{l} \hat{\nabla} \cdot \left( V \vec{l} \right) = V \vec{v} + \vec{l} \left( \vec{l} \cdot \hat{\nabla} V + V \hat{\nabla} \cdot \vec{l} \right) \\
- M^{-1} \vec{l} \cdot \hat{\nabla} \vec{g}^\mu &=& \vec{l} \cdot \hat{\nabla} \left( V \vec{l} l^\mu \right) = V \vec{v} l^\mu + \vec{l} \left( l^\mu \vec{l} \cdot \hat{\nabla} V + V \hvec{e}^\mu \cdot \vec{v} \right),
\eea
\end{subequations}
and that $l^\mu \mv{\Omega}_\mu = \mv{\Omega}(\vec{l}) = 0$, we find
\bea
M^{-1} \tensor{R}(\vec{l}) = 2 \vec{v} \vec{l} \cdot \hat{\nabla} V + V \vec{l} \cdot \hat{\nabla} \vec{v} + \vec{v} V \hat{\nabla} \cdot \vec{l} + \vec{l} \vec{l} \cdot \hat{\nabla} \Theta - V \vec{l} \cdot \mv{\Omega}(\vec{v}),
\label{ricci_l}
\eea
where
\beann
\Theta \equiv \hat{\nabla} \cdot (V \vec{l}) = \vec{l} \cdot \hat{\nabla} V + V \hat{\nabla} \cdot \vec{l} 
\eeann
is the expansion parameter for the vector $V \vec{l}$.  Taking the inner product of \eq{ricci_l} with $\vec{l}$ eliminates all but one term:
\bea
\vec{l} \cdot \tensor{R}(\vec{l}) = M V \vec{l} \cdot \left( \vec{l} \cdot \hat{\nabla} \vec{v} \right) = - M V \vec{v}^2 ,
\label{lRl}
\eea
where the last expression follows from the vanishing of $\vec{v} \cdot \vec{l}$.  When $\vec{v}$ vanishes, $\vec{l}$ is a null geodesic, and the Ricci tensor satisfies Einstein's equations.  (Note that any vector $\vec{v} \propto \vec{l}$ also makes \eq{lRl} vanish, but we can transform the proportionality to zero simply by redefining the time coordinate \cite{nakahara}.)  From here on, we let
\beann
\vec{v} = 0 ,
\eeann
even when considering Einstein's equations in the presence of nonvanishing energy-momentum.

Plugging the geodesic condition into \eq{ricci_l} yields an eigenvalue equation:
\bea
\tensor{R}(\vec{l}) = \vec{l} M \vec{l} \cdot \hat{\nabla} \Theta = \vec{l} M \left\{ (\vec{l} \cdot \hat{\nabla} )^2 V + (\vec{l} \cdot \hat{\nabla} V) \hat{\nabla} \cdot \vec{l} + V \vec{l} \cdot \hat{\nabla} (\hat{\nabla}\cdot \vec{l}) \right\} ,
\label{ricc_eigen}
\eea
so that the vacuum condition is $\vec{l} \cdot \hat{\nabla} \Theta = 0$.  The connection bivector becomes
\bea
\mv{\Omega}_\mu = - M \hvec{e}^\lambda \wedge \hat{\nabla}_\lambda \left( V \vec{l} l_\mu \right).
\eea
Using this knowledge, we determine that $\vec{l}$ is a principal null direction by forming the quantity
\beann
\vec{l} \cdot \tensor{C} (\vec{a} \wedge \vec{l}) = a^\mu \vec{l} \cdot \left\{ \tensor{R}(\vec{g}_\mu \wedge \vec{l}) - \frac{1}{2} \left( \vec{g}_\mu \wedge \tensor{R}(\vec{l}) + \vec{l} \wedge \tensor{R}(\vec{g}_\mu ) \right) + \frac{1}{6} R \vec{g}_\mu \wedge \vec{l} \right\}\eeann
where
\beann
&{}&\vec{l} \cdot \tensor{R} (\vec{g}_\mu \wedge \vec{l}) = M \vec{l} l_\mu \left(\vec{l} \cdot \hat{\nabla} \right)^2 V \\
&{}&\vec{l} \cdot \left( \vec{g}_\mu \wedge \tensor{R}(\vec{l}) + \vec{l} \wedge \tensor{R}(\vec{g}_\mu ) \right) = \vec{l} \cdot \vec{g}_\mu \tensor{R}(\vec{l}) - \vec{l} \vec{l} \cdot \tensor{R}(\vec{g}_\mu) = 0
\eeann
(using the symmetry of the Ricci tensor $\vec{l} \cdot \tensor{R}(\vec{g}_\mu) = \vec{g}_\mu \cdot \tensor{R}(\vec{l})$).  What remains is
\bea
\vec{l} \cdot \tensor{C} (\vec{g}_\mu \wedge \vec{l}) &=& \left\{ M \left(\vec{l} \cdot \hat{\nabla} \right)^2 V + \frac{1}{6} R \right\} \vec{l} \vec{l}\cdot \hvec{e}_\mu \equiv H \vec{l} l_\mu \nonumber \\
\to \quad \vec{l} \cdot \tensor{C} (\vec{a} \wedge \vec{l}) &=& H \vec{l} \vec{l} \cdot \vec{a}
\label{weyl_eigen_kerr}
\eea
which is an eigenvalue equation.  Obviously, in a vacuum the Ricci scalar vanishes, but this does not change the essence of the Weyl tensor.  Along the same lines is the expression
\bea
\vec{l} \wedge \tensor{C}(\vec{a} \wedge \vec{l}) = \vec{l} \wedge \tensor{R}(\vec{a} \wedge \vec{l}) = M \vec{l}\cdot\vec{a} \left( \vec{l} \cdot \hat{\nabla} V \right) \vec{l} \wedge \hat{\nabla} \wedge \vec{l} .
\label{weyl_wedge_lambda}
\eea
From this development, we see that $\vec{l} \cdot (\vec{a} \wedge \vec{l}) = \vec{l} \vec{l}\cdot \vec{a}$ (and hence $\vec{l}$) is a principal null direction, which we will examine in more detail in the subsequent section.  We will confine ourselves to the vacuum case from here on.

\subsection{Kerr Geometry}

To recap, we have found that $\vec{l}$ of the Kerr-Schild ansatz is a principal null direction.  From it, we define the bivector
\bea
\mv{e} = \vec{l} \wedge \hvec{e}_0 ,
\eea
and write
\bea
\vec{l} = ( 1 + \mv{e} )\hvec{e}_0 
\eea
so that $l_0 = \vec{l} \cdot \vec{g}_0 = 1$ and $\mv{e}^2 = 1$.  From this null vector we define the tetrad paravector
\bea
\pv{l} = \frac{1}{2} \vec{l} \hvec{e}_0 = \frac{1}{2} ( 1 + \mv{e} )
\eea
and select an arbitrary unit vector that satisfies the properties
\beann
\hvec{u}^2 = -1 \ , \quad \hvec{u} \cdot \vec{l} = 0 \ , \quad \hvec{u} \cdot \hvec{e}_0 = 0
\eeann
to generate the rest of the tetrad.  More important to us for the moment is the bivector basis from \eq{bivbasis}, which we write as
\bea
\sqrt{2} \mv{Z}_0 = \hvec{u} \wedge \vec{l} \quad \mv{Z}_1 = \mv{e} \quad \sqrt{2} \mv{Z}_2 = \hvec{u} \wedge \left( \hvec{e}_0 \vec{l} \hvec{e}_0 \right).
\eea
From these, we can form the $\Psi_a$ as in Section~\ref{geomclass}.  The first two are identically zero, following from \eq{weyl_eigen_kerr} and \eq{weyl_wedge_lambda}:
\beann
\Psi_0 &=& %- \grade{\mv{Z}_0 \tensor{C} (\mv{Z}_0)}_S = 
\frac{1}{2} \left\{ (\hvec{u} \wedge \vec{l} ) \cdot \tensor{C}(\hvec{u} \wedge \vec{l}) + (\hvec{u} \wedge \vec{l} ) \wedge \tensor{C}(\hvec{u} \wedge \vec{l}) \right\} = \frac{1}{2} \left\{ \hvec{u} \cdot \left( \vec{l} \cdot \tensor{C}(\hvec{u} \wedge \vec{l} ) \right) + \hvec{u} \wedge \left( \vec{l} \wedge \tensor{C}(\hvec{u} \wedge \vec{l} ) \right) \right\} = 0 \\
\Psi_1 &=& %- \grade{\mv{Z}_1 \tensor{C} (\mv{Z}_0)}_S = 
\frac{1}{2} \left\{ (\hvec{e}_0 \wedge \vec{l} ) \cdot \tensor{C}(\hvec{u} \wedge \vec{l}) + (\hvec{e}_0 \wedge \vec{l} ) \wedge \tensor{C}(\hvec{u} \wedge \vec{l}) \right\} = \frac{1}{2} \left\{ \hvec{e}_0 \cdot \left( \vec{l} \cdot \tensor{C}(\hvec{u} \wedge \vec{l} ) \right) + \hvec{e}_0 \wedge \left( \vec{l} \wedge \tensor{C}(\hvec{u} \wedge \vec{l} ) \right) \right\} = 0 ,
\eeann
since $\hvec{u} \cdot \vec{l} = 0$ by definition.  This demonstrates that any Kerr-Schild solution is algebraically special.  The actual Petrov type, however, depends on other choices, which we will make now.

To obtain a black-hole solution, we require that $V$ and $\mv{e}$ be functions of the spatial coordinates only.  With this criterion, we will reduce the field equations to expressions in the three dimensional subalgebra $\Cl^+_{1,3}$.  Since $\hat{\nabla}_0 \mv{A} = 0$, we define the Cartesian 3-gradient $\grad \equiv \hvec{e}^k \hat{\nabla}_k$ and its bivector equivalent
\bea
\vec{\nabla} \equiv \hvec{e}_0 \wedge \grad = \hvec{e}_0 \grad .
\eea
Thus, the four-dimensional gradient can be reduced to
\bea
\hat{\nabla} = \vec{g}^k \hat{\nabla}_k = \grad - M V \vec{l} \mv{e} \cdot \vec{\nabla} ,
\eea
from which we obtain $\vec{l} \cdot \hat{\nabla} = \mv{e} \cdot \vec{\nabla}$.

Following \cite{schiffer}, we now obtain the field equations; since $M$ is an arbitrary constant, each coefficient of a power of $M$ must vanish separately.  The Ricci tensor is
\beann
\tensor{R}_\nu = \vec{g}^\mu \cdot \tensor{R} (\vec{g}_\mu \wedge \vec{g}_\nu ) = \vec{g}^\mu \cdot \left( \hat{\nabla}_\mu \mv{\Omega}_\nu - \hat{\nabla}_\nu \mv{\Omega}_\mu + \mv{\Omega}_\mu \times \mv{\Omega}_\nu \right) ;
\eeann
each $\mv{\Omega}$ contributes a factor of $M$, while $\vec{g}^\mu = \hvec{e}^\mu - M V l^\mu \vec{l}$ contains a zero-order and first-order piece.  Therefore, we have the following expressions:
\begin{subequations}
\bea
\mathcal{O}(M) : &\quad& \hvec{e}^\mu \cdot \left( \hat{\nabla}_\mu \mv{\Omega}_\nu - \hat{\nabla}_\nu \mv{\Omega}_\mu \right) = \grad \cdot \mv{\Omega}_\nu - \hat{\nabla}_\nu \left( \hvec{e}^\mu \cdot \mv{\Omega}_\mu \right) \\
\mathcal{O}(M^2) : &\quad& - M V l^\mu \vec{l} \cdot \left( \hat{\nabla}_\mu \mv{\Omega}_\nu - \hat{\nabla}_\nu \mv{\Omega}_\mu \right) + \hvec{e}^\mu \cdot \left( \mv{\Omega}_\mu \times \mv{\Omega}_\nu \right) \\
\mathcal{O}(M^3) : &\quad& - M V l^\mu \vec{l} \cdot \left( \mv{\Omega}_\mu \times \mv{\Omega}_\nu \right) = - M V \vec{l} \cdot \left( \mv{\Omega}(\vec{l}) \times \mv{\Omega}_\nu \right) = 0 .
\eea
\end{subequations}
As the reader can easily tell, we have reduced the complexity of the problem significantly:  the first-order equation, which we will treat momentarily, is straightforward to find, while the third-order expression is an identity.  The remaining equation is satisfied by the solution to the first-order equation; we will not prove this here, since it does not involve anything new.

\subsubsection*{$\mathcal{O}(M)$ Field Equations}

After some work we find that
\beann
M^{-1} \grad \cdot \mv{\Omega}_\nu = \grad{}^2 \left( V \vec{l} l_\nu \right) + \grad \left( l_\nu \Theta \right),
\eeann
where $\Theta = \hat{\nabla} \cdot ( V \vec{l} ) = \mv{e} \cdot \vec{\nabla} V + V \vec{\nabla} \cdot \mv{e}$ as before, and we have used the vanishing of $\vec{l} \cdot \hat{\nabla} \Theta = \mv{e} \cdot \vec{\nabla} \Theta$ from \eq{ricc_eigen}.  We already know that
\beann
- \hvec{e}^\mu \cdot \mv{\Omega}_\mu = - \vec{g}^\mu \cdot \mv{\Omega}_\mu = M \vec{l} \Theta ,
\eeann
so that the first-order field equations are
\bea
\grad{}^2 \left( V \vec{l} l_\nu \right) + \grad \left( l_\nu \Theta \right) + \hat{\nabla}_\nu \left( \vec{l} \Theta \right) = 0 .
\label{OM}
\eea

When $\nu = 0$, \eq{OM} reduces to 
\bea
\grad{}^2 \left( V \vec{l} \right) + \grad \Theta = \vec{\nabla}^2 \Big\{ V \left( \hvec{e}_0 + \mv{e} \cdot \hvec{e}_0 \right)\Big\} + \hvec{e}_0 \cdot \vec{\nabla} \Theta = 0 ;
\eea
multiplying by $\hvec{e}_0$ splits the equation into scalar and bivector pieces:
\bea
\vec{\nabla}^2 V = 0 \qquad \text{and} \qquad
\vec{\nabla}^2 \left( V \mv{e} \right) + \vec{\nabla} \Theta = 0 .
\label{nu0_OM}
\eea
For the case when $\nu = k$ (for $k = 1,2,3$), we have
\bea
\grad{}^2 \left( V \vec{l} l_k \right) + \grad \left( l_k \Theta \right) + \hat{\nabla}_k \left( \vec{l} \Theta \right) = \left\{ \grad{}^2 \left( V \hvec{e}_0 l_k \right) + \hvec{e}_0 \hat{\nabla}_k \Theta \right\} + \hvec{e}^j \grad{}^2 \left( V l_j l_k \right) + \grad \left( l_k \Theta \right) + \hat{\nabla}_k \left( \mv{e} \Theta \right) \cdot \hvec{e}_0 ;
\eea
the term in braces vanishes due to \eq{nu0_OM}.  Expanding the Laplacian yields
\bea
\grad{}^2 \left( V l_j l_k \right) = l_j \grad{}^2 \left( V l_k \right) + l_k \grad{}^2 \left( V l_j \right) + 2 V \left( \grad l_j \right) \cdot \left( \grad l_k \right) = l_j \hat{\nabla}_k \Theta + l_k \hat{\nabla}_j \Theta + 2 V \left( \grad l_j \right) \cdot \left( \grad l_k \right) ,
\eea
so that the field equations are
\bea
2 V \hvec{e}^j \left( \grad l_j \right) \cdot \left( \grad l_k \right) + \Theta \left\{ \grad l_k + \left( \hat{\nabla}_k \mv{e} \right) \cdot \hvec{e}_0 \right\} = 0
\eea
or, in bivector language,
\bea
2 V \left\{ \vec{\nabla} \left( \hvec{e}_0 \wedge \hvec{e}_k \right) \cdot \mv{e} \right\} \cdot \vec{\nabla} \mv{e} + \Theta \left[ \vec{\nabla} \left\{ \left( \hvec{e}_0 \wedge \hvec{e}_k \right) \cdot \mv{e} \right\} + \left( \hvec{e}_0 \wedge \hvec{e}_k \right) \cdot \vec{\nabla} \mv{e} \right] = 0 .
\eea
Our task is now to find a reasonable expression for the derivatives of $\mv{e}$; once again, we follow \cite{schiffer}.

For some arbitrary vector $\vec{a}$ we define the bivector $\mv{a} = \hvec{e}_0 \wedge \vec{a}$, and from it create the operator
\bea
\op{M}(\mv{a}) \equiv \mv{a} \cdot \vec{\nabla} \mv{e} ,
\eea
whose adjoint is
\bea
\adj{M}(\mv{a}) = \grave{\vec{\nabla}} \left( \mv{a} \cdot \grave{\mv{e}} \right) = \vec{\nabla} \left( \mv{a} \cdot \mv{e} \right) - \mv{\partial_b} \left( (\mv{b} \cdot \vec{\nabla} \mv{a} ) \cdot \mv{e} \right) ;
\eea
we have used the accents to show that $\mv{a}$ is not differentiated.  From these definitions, we see that the field equations can be rewritten as
\bea
\frac{2 V}{\Theta} \op{M}\adj{M}(\mv{a}) = \op{M}(\mv{a}) + \adj{M}(\mv{a}),
\eea
and we note that $\op{M}(\mv{e}) = \adj{M}(\mv{e}) = 0$.  Due to the latter property, we make the following ansatz:
\bea
\op{M}(\mv{a}) = \mv{a} - \op{U}(\mv{a})
\eea
where $\op{U}$ is a unitary operator, which obeys $\op{U}(\mv{e}) = \adj{U}(\mv{e}) = \mv{e}$.  A unitary operator that preserves a certain spatial direction can be written as a rotation:
\bea
\op{U} (\mv{a}) = \text{e}^{i \mv{e} \psi/2} \mv{a} \text{e}^{- i \mv{e} \psi/2}
\eea
which means that, after expanding the exponentials, we find
\bea
\mv{a} \cdot \vec{\nabla} \mv{e} = \op{M}(\mv{a}) = \alpha ( \mv{a} \times \mv{e} ) \mv{e} + \beta i (\mv{a} \times \mv{e} ) ,
\label{dirderiv_e}
\eea
where
\bea
\alpha = \frac{\Theta}{2 V} (1 - \cos \psi ) \quad \text{and} \quad \beta = \frac{\Theta}{2 V} \sin \psi .
\eea
(Recall that the symbol $\times$ represents the \emph{commutator product}; if we were using the vector cross product, the pseudoscalar element $i$ would not be present.)  Removing the arbitrary bivector $\mv{a}$ gives us the divergence and curl:% \footnote{There is a sign difference in $\beta$ between our technique and that of \cite{schiffer} or \cite{ks2}, so that our $\gamma = \alpha + i \beta$ is the complex conjugate of the others.  This can be erased by taking the opposite sign for $\psi$.  We choose this rotation direction to obtain a simpler formula for the connection.}:
\bea
\vec{\nabla} \mv{e} &=& \mv{\partial_a} \mv{a} \cdot \vec{\nabla} \mv{e} = 2 \alpha + 2 \beta i \mv{e} \nonumber \\
\to \  \vec{\nabla} \cdot \mv{e} &=& 2 \alpha \quad \text{and} \quad \vec{\nabla} \times \mv{e} = 2 i \beta \mv{e}.
\eea

Following \cite{ks2} again, we form the Laplacian of $\mv{e}$ via \eq{dirderiv_e} and by direct calculation (in order of mention):
\begin{subequations}
\bea
\vec{\nabla}^2 \mv{e} &=& \vec{\nabla} \cdot \vec{\nabla} \mv{e} = \vec{\nabla} \alpha - \left( \mv{e} \cdot \vec{\nabla} \alpha \right) \mv{e} - 2 \left( \alpha^2 + \beta^2 \right) \mv{e} + i \mv{e} \times \vec{\nabla} \beta \\
\vec{\nabla}^2 \mv{e} &=& \grade{ \vec{\nabla} (\vec{\nabla}\mv{e})}_2 = 2 \left( \vec{\nabla} \alpha - i \vec{\nabla} \times (\beta \mv{e}) \right) = 2 \left( \vec{\nabla} \alpha + i \mv{e} \times \vec{\nabla} \beta - 2 \beta^2 \mv{e} \right)
\eea
\label{laplacian_e}
\end{subequations}
where we have used the identity $\vec{\nabla} \times (\beta \mv{e}) = \beta \vec{\nabla} \times \mv{e} - \mv{e} \times \vec{\nabla} \beta$.  If we further note that
\beann
\vec{\nabla} \cdot (\beta\mv{e}) = 0 = \beta \vec{\nabla} \cdot \mv{e} + \mv{e} \cdot \vec{\nabla} \beta = 2 \alpha \beta + \mv{e} \cdot \vec{\nabla} \beta
\eeann
and combine the Laplacian expressions, we obtain
\bea
\begin{array}{lll}
\vec{\nabla} \alpha = (\beta^2 - \alpha^2 ) \mv{e} + i \mv{e} \times \vec{\nabla} \beta & \quad & \vec{\nabla} \beta = - 2 \alpha \beta \mv{e} - i \mv{e} \times \vec{\nabla} \alpha \\
\mv{e} \cdot \vec{\nabla} \alpha = \beta^2 - \alpha^2 & \quad & \mv{e} \cdot \vec{\nabla} \beta = - 2 \alpha \beta .
\end{array}
\label{alpha_beta_derivs}
\eea
If we let $\gamma = \alpha + i \beta$, we can collect the directional derivatives into a single equation:
\bea
\mv{e} \cdot \vec{\nabla} \gamma = - \gamma^2 ,
\eea
while
\bea
\vec{\nabla} \gamma = - \gamma^2 \mv{e} + \mv{e} \times \vec{\nabla} \gamma .
\eea
From this, a little more work shows that
\bea
\vec{\nabla}^2 \gamma = 0 ,
\eea
and
\bea
\left( \vec{\nabla} \gamma^{-1} \right)^2 = 1 ;
\eea
thus, $\gamma$ is harmonic, and its inverse satisfies the eikonal equation.  

Since both $V$ and $\alpha$ are harmonic, it is reasonable to ask whether they are equal; it turns out this guess is consistent, by evaluating \eq{nu0_OM}.  Noting that $V=\alpha$ yields
\bea
\Theta = \mv{e} \cdot \vec{\nabla} \alpha + \alpha \vec{\nabla} \cdot \mv{e} = \alpha^2 + \beta^2 = \gamma \gamma^\dagger ,
\eea
we work out the Laplacian using \eq{laplacian_e} and \eq{alpha_beta_derivs} to give us
\bea
\vec{\nabla}^2 ( \alpha \mv{e} ) = \alpha \vec{\nabla}^2 \mv{e} + 2 (\vec{\nabla} \alpha ) \cdot \vec{\nabla} \mv{e} = 2 ( \alpha \vec{\nabla} \alpha + \beta \vec{\nabla} \beta ) = \vec{\nabla} (\alpha^2 + \beta^2) .
\eea
With all of this in hand, we find we can write $\mv{e}$ in terms of $\gamma^{-1}$:
\bea
\mv{e} = \frac{ \vec{\nabla} \gamma^{-1} + \vec{\nabla} (\gamma^{-1})^\dagger - \vec{\nabla} \gamma^{-1} \times \vec{\nabla} (\gamma^{-1})^\dagger }{1 + (\vec{\nabla} \gamma^{-1} ) \cdot (\vec{\nabla} (\gamma^{-1})^\dagger ) }.
\eea

The Schwarzschild solution is found by letting $\gamma = V = 1/r$, so that $\beta = 0$.  The Kerr solution is a little more complicated:
\bea
\gamma^{-2} &=& x^2 + y^2 + (z - i a)^2 \nonumber \\
V &=& \grade{\gamma}_0 = \frac{\rho^3}{\rho^4 + a^2 z^2} \nonumber \\
\mv{e} &=& \frac{1}{a^2 + \rho^2} \left( (\rho x + a y) \vec{\sigma}_1 + (\rho y - a x) \vec{\sigma}_2 \right) + \frac{z}{\rho} \vec{\sigma}_3
\eea
where $\rho$ is implicitly defined by
\bea
\rho^4 - \rho^2 (r^2 - a^2) - a^2 z^2 = 0\ ; \qquad r^2 \equiv x^2 + y^2 + z^2 .
\eea
This value for $\gamma$ is picked to ensure that the system is axisymmetric (about the $z$-axis); the parameter $a$ is related to the angular momentum of the black hole.  %Still needs work!!!!

\section{Further Directions}

We have by no means exhausted the task of translating solution techniques into the geometric algebra language; indeed, we believe that further insights into the geometrical structure of general relativity can be obtained through this process, and perhaps more techniques can be discovered.  In addition, we point to a paper by Sobczyk \cite{killing}, which discusses the embedding of the spacetime of general relativity in a pseudo-Euclidean space of higher dimension, utilizing geometric algebra throughout.

One area of interest is the extension of these techniques to the space of Lorentz spinors (also known as 2-spinors \cite{penrose_rindler1}).  The work in Section~\ref{null_tetrad} is a hint toward this, as are the papers by Jones and Baylis \cite{jones}, and Lasenby, Doran, and Gull \cite{2spinors}.  Spinor techniques lead to an adaptation of the Newman-Penrose formalism \cite{newman-penrose}, which has been useful for obtaining many exact solutions.

The viewpoint in the first section of this paper, in which an
orthonormal frame of vectors is considered to be fundamental (as
opposed to taking a coordinate metric as fundamental) gives a somewhat different ontological status of gravitation, compared with the usual
interpretation of general relativity.  Specifically, this construction can be interpreted as a flat spacetime theory, in which gravitation is a modification of the geometric relationships of an underlying flat spacetime; mathematically, this is locally equivalent to general relativity, which is commonly viewed as
a theory of intrinsically curved spacetime.  We also emphasize that these
two viewpoints are physically indistinguishable, as geodesics are the same in both. %The physical difference is essentially whether a curved or a flat spacetime metric is taken as the foundation for building the mathematical description of gravitation.

If gravitation is viewed as occurring in a Minkowski space arena, it
becomes conceptually closer to the other fundamental forces. Indeed,
the same covariant derivative can also be obtained in a flat
background spacetime by enforcing local gauge invariance under the
gauge group $\mathrm{SL}(2,\mathbb{C})$, the proper Lorentz group \cite{GTG,SCGT}. %A different attempt, based on the gauge group of translations, leads to the so-called ``teleparallel'' formulation of gravitation.  Its gravitational field corresponds to spacetime torsion instead of curvature, but it has been demonstrated that this theory is physically equivalent of general relativity \cite{maluf94,andrade97} for scalar matter sources.
Many fundamental questions related to these
gauge formulations remain unanswered, but the existence of these
theories stands as an interesting contrast to the usual approaches to
gravitation. As the expansion of the Universe is accelerating and
general relativity remains unquantized, alternate approaches may prove
fruitful. The formalism introduced in this paper lends itself quite
naturally to this class of theories.

\begin{acknowledgments}

Harry Zapolsky provided perceptive comments about an earlier edition of this paper.  We thank David Hestenes for his bibliographic suggestions and clarifications of much of his own research.  This work has been partially supported by NASA's Space Astrophysics Research and Analysis program through grant NAG5-10110. AK is a Cottrell Scholar of the Research Corporation.

\end{acknowledgments}

\begin{appendix}

\section{Review of Geometric (Clifford) Algebra}
\label{intro_CA}

Many excellent introductions to geometric (or Clifford) algebra exist (for example \cite{imag_numbs,intro_GA,classical,lounesto}), as well as more detailed mathematical treatises (see \cite{CA2GC} and the various conference proceedings listed in the bibliography), so we will make no attempt to be exhaustive.  Rather, this appendix and the one following are intended to cover the basics in a small amount of space, to get the interested reader started.

\subsection{Geometric Algebra}

The geometric (or Clifford) product is associative and distributive over addition:
\bea
	\vec{a}(\vec{bc}) &=& (\vec{ab})\vec{c} \nonumber \\
	\vec{a}(\vec{b}+\vec{c}) &=& \vec{ab}+\vec{ac}
\eea
and provides a measure of the length of a vector:
\bea
	\vec{a}^2 = \vec{aa} = |\vec{a}|^2 .
\eea
Thus, if we define a vector $\vec{a} = \vec{b}+\vec{c}$, using the axioms above, we obtain
\bea
	\vec{a}^2 = (\vec{b}+\vec{c})(\vec{b}+\vec{c}) = \vec{bb} + \vec{cc} + \vec{bc}+\vec{cb} = \vec{b}^2 + \vec{c}^2 + \vec{bc}+\vec{cb} .
\eea
Since the left side is a scalar, the right side must be as well, so we define the inner product of vectors to be the symmetric sum
\bea
	\vec{b}\cdot\vec{c} = \frac{1}{2} \left( \vec{bc}+\vec{cb} \right) = \vec{c} \cdot \vec{b}
\eea
which vanishes when $\vec{b}$ is orthogonal to $\vec{c}$.  

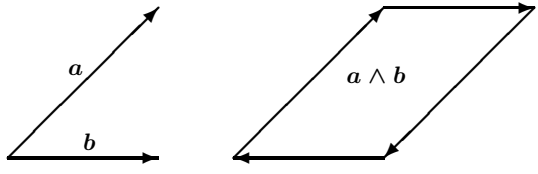
\begin{figure}[t]
\begin{picture}(7.5,3)(-0.5,-0.5)\thicklines
        \put(0,0){\vector(1,1){2}} \put(0.8,1.1){$\bs{a}$}
        \put(0,0){\vector(1,0){2}} \put(1,0.1){$\bs{b}$}
        \put(3,0){\vector(1,1){2}}
        \put(5,0){\vector(-1,0){2}} \put(4.5,1){$\bs{a \wedge b}$}
        \put(5,2){\vector(1,0){2}}
        \put(7,2){\vector(-1,-1){2}}
\end{picture}
\caption{Vectors and bivectors illustrating the outer product.}
\label{vec_bivec}
\end{figure}

The outer product is then the antisymmetric piece of the geometric product:
\bea
	\vec{b} \wedge \vec{c} = \frac{1}{2} \left( \vec{bc}-\vec{cb} \right) = -\vec{c} \wedge \vec{b};
\eea
it is identical to the outer product in differential geometry.  Its value is neither a vector nor a scalar, but an oriented plane segment called a \emph{bivector}, whose sides are defined by the vectors making it up (see \fig{vec_bivec}).  The outer product is associative---
\bea
	\vec{a}\wedge(\vec{b}\wedge\vec{c}) = (\vec{a}\wedge\vec{b})\wedge\vec{c}
\eea
---and vanishes if the vectors are linearly-dependent.

Thus, the geometric product of two vectors is the sum of the inner product and outer product---that is, the sum of a scalar and a bivector:
\bea
	\vec{ab} = \vec{a}\cdot\vec{b} + \vec{a}\wedge\vec{b} = \vec{b}\cdot\vec{a} - \vec{b}\wedge\vec{a} \neq \pm \vec{ba}.
\eea
When we add objects of different types---called \emph{grades}---together, we have a general object called a \emph{multivector}.  We denote a general multivector as the sum of its grades:
\bea
	\mv{A} = \grade{A}_0 + \grade{A}_1 + \grade{A}_2 + \ldots = \sum_{\gamma} \grade{A}_\gamma
\eea
where $\grade{A}_0$ is the scalar part, $\grade{A}_1$ is the vector part, and so on.  The grade subscript $\gamma$ is an indication of how many vectors are needed to construct the multivector via the outer product.  

Therefore, we write
\bea
	\vec{a}\cdot\vec{b} = \grade{\vec{ab}}_0 \quad \mbox{and}\quad \vec{a}\wedge\vec{b} = \grade{\vec{ab}}_2
\eea
for the product of two vectors.  We can extend the geometric product to cover arbitrary multivectors, so that
\bea
	\vec{a}\grade{\mv{A}}_r = \vec{a}\cdot\grade{\mv{A}}_r + \vec{a}\wedge\grade{\mv{A}}_r
\eea
where
\bea
	\vec{a}\cdot\grade{\mv{A}}_r = \frac{1}{2} \left( \vec{a}\mv{A} - (-1)^r \mv{A}\vec{a} \right)
\eea
has a grade of $r - 1$ (and vanishes if $r=0$), while 
\bea
	\vec{a}\wedge\grade{\mv{A}}_r = \frac{1}{2} \left( \vec{a}\mv{A} + (-1)^r \mv{A}\vec{a} \right)
\eea
has a grade of $r+1$.

The geometric product also covers multiplication of arbitrary multivectors.  We are still most interested in the inner and outer products:
\bea
	\grade{\mv{A}}_r \cdot \grade{\mv{B}}_s = \grade{\mv{AB}}_{|r-s|} \nonumber \\
	\grade{\mv{A}}_r \wedge \grade{\mv{B}}_s = \grade{\mv{AB}}_{r+s},
\eea
but we also define the commutator product:
\bea
	\mv{A}\times\mv{B} = \frac{1}{2} (\mv{AB}-\mv{BA})
\eea
which should not be confused with the vector cross product.  It obeys the Jacobi identity
\bea
\mv{A}\times(\mv{B}\times\mv{C}) + \mv{B}\times(\mv{C}\times\mv{A}) + \mv{C}\times(\mv{A}\times\mv{B}) = 0
\eea
and is ``distributive'' over the geometric product:
\bea
\mv{A}\times(\mv{BC}) = (\mv{A}\times\mv{B})\mv{C} + \mv{B}(\mv{A}\times\mv{C}).
\eea
The commutator product is particularly useful with bivectors, since it preserves the grade of the object it multiplies:
\bea
	\grade{\mv{B}}_2 \times \grade{\mv{C}}_r = \grade{\mv{B}\times\mv{C}}_r.
\eea

One final operation of note is the \emph{reverse}, which is defined as
\bea
\widetilde{\mv{AB}} = \tilde{\mv{B}}\tilde{\mv{A}}
\eea
for all multivectors, while
\bea
\tilde{\vec{a}} = \vec{a}
\eea
for all vectors.  Thus, bivectors change sign under reversion, which geometrically means that their orientation is reversed.

\subsection{Spacetime Algebra}

Up to this point, we have not specified a dimension for our space.  Since this paper deals with spacetime, we need a space of four dimensions, with a Lorentz signature (\emph{i.e.}, the metric's trace is $-2$, for consistency with spinorial methods).  In geometric algebra, this space is denoted by $\Cl_{1,3}$ \cite{lounesto}, and it is frequently useful to define (locally, at least) a set of basis vectors $\{ \vec{\gamma}_\mu \}$, which satisfy
\bea
	\vec{\gamma}_\mu \cdot \vec{\gamma}_\nu = \eta_{\mu\nu}
\eea
where $\eta = \mathrm{diag}(+1,-1,-1,-1)$ is the metric for a flat Lorentzian spacetime.  (Other frames of vectors can be defined; a more precise definition of frames is laid out in Section \ref{frames}, as well as in \cite{SCGT}.)  Due to their linear independence, we know that
\bea
	\vec{\gamma}_\kappa \wedge \vec{\gamma}_\lambda \wedge \vec{\gamma}_\mu \wedge \vec{\gamma}_\nu \wedge\vec{\gamma}_\xi = 0
\eea
since there do not exist five linearly-independent vectors in the spacetime.  In addition,
\bea
	\vec{\gamma}_\kappa \wedge \vec{\gamma}_\lambda \wedge \vec{\gamma}_\mu \wedge \vec{\gamma}_\nu = \epsilon_{\kappa\lambda\mu\nu} i
\eea
where $\epsilon_{\kappa\lambda\mu\nu}$ is the completely antisymmetric Levi-Civita symbol, while
\bea
	i \equiv \vec{\gamma}_0 \wedge \vec{\gamma}_1 \wedge \vec{\gamma}_2 \wedge \vec{\gamma}_3 = \vec{\gamma}_0 \vec{\gamma}_1 \vec{\gamma}_2 \vec{\gamma}_3
\eea
is the unit tetravector, or pseudoscalar, for spacetime.  It has the special properties
\bea
	i^2 = -1 \quad \mbox{and} \quad i \grade{\mv{A}}_r = (-1)^r\grade{\mv{A}}_r i
\eea
for grade-$r$ multivectors.  In addition, the pseudoscalar can be used to define the inner product from the outer product \cite{CA2GC}:
\bea
\vec{a} \cdot \mv{B} = - \vec{a} \wedge \left(\mv{B}i\right) i 
\label{duality}
\eea
for any vector $\vec{a}$ and generic multivector $\mv{B}$; the more general version of this relationship demonstrates that the existence of a pseudoscalar is equivalent to the existence of a metric.

The reader may have noticed a strong similarity between the algebra of the basis vectors $\vec{\gamma}_\mu$, and the algebra of the Dirac matrices $\gamma_\mu$.  Indeed, the Dirac matrices are a representation of the Clifford algebra $\Cl_{1,3}$, which describes spacetime.  For more information about matrix representations, please see \cite{lounesto}.

\subsection{The Even Subalgebra $\Cl^+_{1,3}$}
\label{even_subalgebra}

The even multivectors (scalars, bivectors, and pseudoscalars) form a subalgebra, which we call the \emph{even subalgebra}, and denote with $\Cl^+_{1,3}$.  As an algebra, it is isomorphic to the three-dimensional geometric algebra $\Cl_3$, and it is possible to perform many spacetime calculations in this restricted space; see \cite{baylis,jones,lounesto} for more information.

Proper, orthochronous Lorentz transformations can be written as \cite{STA}
\bea
\op{\Lambda} (\mv{A}) = \mv{U A}\tilde{\mv{U}} = \mathrm{e}^{\mv{B}/2} \mv{A} \mathrm{e}^{-\mv{B}/2}
\eea
where $\mv{A}$ is an arbitrary multivector, $\mv{B}$ is a bivector, and the exponential $\mv{U} = \exp(\mv{B}/2)$ is defined (as in matrix algebra) by its Taylor series:
\bea
\mathrm{e}^{\mv{B}} = 1 + \mv{B} + \frac{1}{2} \mv{B}^2 + \frac{1}{6} \mv{B}^3 + \ldots .
\eea
The quantity
\bea
\mv{B}^2 = \mv{B}\cdot\mv{B} + \mv{B}\wedge\mv{B}
\eea
is in general a scalar added to a pseudoscalar, since in four dimensions the exterior product of bivectors cannot be assumed to vanish.  Thus, this exponential is always a member of the spin group $\mathrm{Spin}_+(1,3) \simeq \mathrm{SL}(2,\mathbb{C})$, whose Lie algebra is isomorphic to the even subalgebra $\Cl^+_{1,3}$ \cite{lounesto}.  This subalgebra consists of the scalars, bivectors, and pseudoscalars:
\bea
\pv{\psi} = \grade{\pv{\psi}}_0 + \grade{\pv{\psi}}_2 + \grade{\pv{\psi}}_4
\eea
and is closed under the geometric product; all even multivectors commute with the pseudoscalar element as well.  We call any object in this subalgebra a (Dirac) \emph{spinor}, following the convention of Hestenes \cite{realspinors,intro_GA}.

The reversion operation allows us to split the algebra into (complex) scalar and bivector pieces:
\bea
\grade{\pv{\psi}}_S &\equiv& \frac{1}{2} \left( \pv{\psi} + \tilde{\pv{\psi}} \right) = \grade{\pv{\psi}}_0 + \grade{\pv{\psi}}_4 \nonumber \\
\grade{\pv{\psi}}_B &\equiv& \frac{1}{2} \left( \pv{\psi} - \tilde{\pv{\psi}} \right) = \grade{\pv{\psi}}_2 ;
\eea
combining these two operations allows us to decompose any spinor into members of four subspaces:
\bea
\pv{\psi} = \grade{\pv{\psi}}_{\Re S} + \grade{\pv{\psi}}_{\Re B} + \grade{\pv{\psi}}_{\Im B} + \grade{\pv{\psi}}_{\Im S},
\eea
where
\bea
\grade{\pv{\psi}}_{\Re S} \equiv \frac{1}{4} \left( \pv{\psi} + \pv{\psi}^\dagger + \tilde{\pv{\psi}} + \tilde{\pv{\psi}}^\dagger \right) 
\eea
is the real scalar part, and so forth for all combinations.

\section{Geometric Calculus and Tensor Algebra}
\label{intro_GC}

The general results of this section are adapted from \cite{CA2GC}.

\subsection{Directional and Vector Derivatives}

Consider an arbitrary function of a vector $\tensor{f}(\vec{a})$.  The rate of change of this function along another vector $\vec{b}$ is the \emph{directional derivative}, which we write as
\bea
\vec{b} \cdot \vec{\partial_a} \tensor{f}(\vec{a}) \equiv \lim_{\epsilon \to 0} \frac{1}{\epsilon} \tensor{f} (\vec{a} + \epsilon \vec{b}) - \tensor{f}(\vec{a});
\label{directional_deriv}
\eea
the notation $\vec{\partial}_a$ indicates that the derivative is with respect to the vector $\vec{a}$.  This derivative is scalar-valued, so it preserves the grade of the object on which it acts; we also note that when $\tensor{f}(\vec{a}) = \vec{a}$, we obtain $\vec{b} \cdot \vec{\partial_a} \vec{a} = \vec{b}$.  The coordinate derivative is the most useful special case of the directional derivative, and the one we use most often in this paper; we find it from letting $\vec{a} = \vec{x}$ and $\vec{b} = \vec{g}_\mu$, for some coordinate frame:
\bea
\partial_\mu \tensor{f}(\vec{x}) \equiv \vec{g}_\mu \cdot \vec{\partial_x} \tensor{f}(\vec{x}) = \vec{g}_\mu \cdot \vec{\partial} \tensor{f}(\vec{x}) .
\eea
As indicated here, we let $\vec{\partial} \equiv \vec{\partial_x}$.  (Equivalently, vectors are frequently defined as directional derivatives.)  The directional derivative is linear, and obeys the Leibniz rule:
\bea
\vec{a}\cdot \vec{\partial} (\mv{A} + \mv{B}) &=& \vec{a}\cdot \vec{\partial} \mv{A} + \vec{a}\cdot \vec{\partial} \mv{B} \nonumber \\
(\vec{a} + \vec{b}) \cdot \vec{\partial} \mv{A} &=& \vec{a}\cdot \vec{\partial} \mv{A} + \vec{b}\cdot \vec{\partial} \mv{A} \nonumber \\
\vec{a}\cdot \vec{\partial} \mv{A B} &=& (\vec{a}\cdot \vec{\partial} \mv{A})\mv{B} + \mv{A} (\vec{a}\cdot \vec{\partial} \mv{B}) .
\eea

We have implicitly defined $\vec{\partial_a}$ in this discussion, but its explicit form requires the introduction of some arbitrary basis $\vec{e}_\mu$, and the use of \eq{directional_deriv}:
\bea
\vec{\partial_a} = \vec{e}^\mu \vec{e}_\mu \cdot \vec{\partial_a} .
\eea
In turn, this allows us to write the gradient
\bea
\vec{\partial} = \vec{\partial_a} \vec{a} \cdot \vec{\partial},
\eea
which we can expand and evaluate using all the previous relations.  The operator $\vec{\partial_a}$ is vector-valued, and therefore does not commute with arbitrary multivectors.  Due to this, we will find the following identities useful:
\bea
\vec{\partial_a} ( \grade{\mv{A}}_r \cdot \vec{a} ) &=& (-1)^{r+1} \vec{\partial_a} ( \vec{a} \cdot \grade{\mv{A}}_r ) = (-1)^{r+1} r \grade{\mv{A}}_r \nonumber \\
\vec{\partial_a} ( \grade{\mv{A}}_r \wedge \vec{a} ) &=& (-1)^r \vec{\partial_a} ( \vec{a} \wedge \grade{\mv{A}}_r ) = (-1)^r (4-r) \grade{\mv{A}}_r \nonumber \\
\vec{\partial_a} ( \grade{\mv{A}}_r \vec{a} ) &=& \vec{\partial_a} ( \grade{\mv{A}}_r \cdot \vec{a} ) + \vec{\partial_a} ( \grade{\mv{A}}_r \wedge \vec{a} ) = (-1)^r (4 - 2 r) \grade{\mv{A}}_r
\eea
where $\mv{A}$ is independent of $\vec{a}$.  Note that these also cover scalars ($r=0$), so that
\bea
\vec{\partial_a} \vec{a} = \vec{\partial_a} \cdot \vec{a} = 4 \qquad \text{and} \qquad \vec{\partial_a} \wedge \vec{a} = 0 .
\eea

\subsection{Tensors}

Tensors are linear functions of one or more vector or (more rarely) multivector variables.  The range of the tensors may contain any multivector, but we will most often deal with tensors that preserve the rank of their argument; this class covers the vierbein fields, the curvature tensors (Riemann, Weyl, and Ricci), and Lorentz transformations; the only real exception we deal with in this paper is the connection bivector, which takes a vector argument and returns a bivector.  (The connection is usually not considered a tensor, since it transforms inhomogeneously under the Lorentz group; however, $\mv{\Omega}(\vec{b})$ is linear in its argument, and so is a tensor in that sense.  See Appendix~\ref{connection_derivation}.)  Since tensors are linear, they obey the important relation
\bea
\tensor{T}(\alpha \mv{A} + \beta \mv{B}) = \alpha \tensor{T}(\mv{A}) + \beta \tensor{T}(\mv{B}),
\eea
where $\alpha$ and $\beta$ are scalars, and $\mv{A}$ and $\mv{B}$ are multivectors.  By the same token, for any scalar $\alpha$, $\tensor{T}(\alpha) = \alpha$.

Now let us consider the class of tensors that map vectors onto vectors:
\beann
\vec{b}' = \op{T}(\vec{b}),
\eeann
where we use an underscore to indicate that grade is preserved.  The components of the tensor with respect to an arbitrary basis $\{ \vec{e}_\mu \}$ are
\bea
T_{\mu\nu} = \vec{e}_\mu \cdot \tensor{T}(\vec{e}_\nu ),
\eea
so that
\beann
\op{T}(\vec{a}) = a^\mu \op{T}(\vec{e}_\mu) = a^\mu T^\lambda{}_\mu \vec{e}_\lambda .
\eeann
The adjoint (or transpose) of a tensor is found from the definition
\bea
\vec{a} \cdot \op{T}(\vec{b}) = \vec{b} \cdot \adj{T}(\vec{a}) \qquad \to \qquad \adj{T}(\vec{b}) = \vec{\partial_a} \vec{a} \cdot \op{T}(\vec{b}),
\eea
where $\vec{\partial_a}$ is defined as in the previous section.  A tensor of this type can be extended to operate on a general multivector via the \emph{outermorphism}:
\bea
\op{T}\Big( \bigwedge\limits^n_{j=1} \vec{a}_j \Big) = \bigwedge\limits^n_{j=1} \op{T}\left(\vec{a}_j\right)
\eea
where
\bea
\bigwedge\limits^n_{j=1} \vec{a}_j \equiv \vec{a}_1 \wedge \vec{a}_2 \wedge \ldots \wedge \vec{a}_n .
\eea
(Note that curvature, although it takes a bivector argument, is \emph{not} an outermorphism:  it cannot be decomposed as $\tensor{R}(\vec{a}\wedge\vec{b}) = \tensor{R}'(\vec{a}) \wedge \tensor{R}'(\vec{b})$.)  Since the pseudoscalar $i$ is unique to a multiplicative scalar, the outermorphism acting on a pseudoscalar must be a defined quantity:
\bea
\op{T}(i) = i \det T  \quad \to \quad \det T = - i \op{T}(i),
\eea
where $\det T$ is the determinant of the tensor.  If the determinant is nonzero, we can write the inverse of the operator as
\bea
\inv{T}(\vec{a}) = \adj{T}(\vec{a} i) (i \det T )^{-1} .
\eea

The \emph{trace} of a tensor of this type is defined as
\bea
\text{tr} T \equiv \vec{\partial_a} \cdot \op{T}(\vec{a}) = \vec{e}^\mu \cdot \op{T}(\vec{e}_\mu) = \vec{e}^\mu \cdot \vec{e}_\lambda T^\lambda{}_\mu = T^\mu{}_\mu ,
\eea
where we have defined a basis for evaluation.  When operating on a tensor of a more general type, this is called the \emph{contraction}:
\bea
\vec{\partial_a} \cdot \tensor{T}(\vec{a},\vec{b},\ldots) = \tensor{S}(\vec{b},\ldots) .
\eea
The \emph{protraction} and \emph{traction} are defined analogously, using the outer and geometric products, respectively:
\bea
\vec{\partial_a} \wedge \tensor{T}(\vec{a},\vec{b},\ldots) \qquad \vec{\partial_a} \tensor{T}(\vec{a},\vec{b},\ldots) = \vec{\partial_a} \cdot \tensor{T}(\vec{a},\vec{b},\ldots) + \vec{\partial_a} \wedge \tensor{T}(\vec{a},\vec{b},\ldots) .
\eea
The main place where these operations are used in this paper is in defining the Ricci tensor,
\bea
\tensor{R}(\vec{a}) = \vec{\partial_b} \tensor{R}(\vec{b} \wedge \vec{a}) = \vec{\partial_b} \cdot \tensor{R}(\vec{b} \wedge \vec{a}) \qquad \to \quad \vec{\partial_b} \wedge \tensor{R}(\vec{b} \wedge \vec{a}) = 0 ,
\eea
and in noting that the Weyl tensor is ``tractionless'':
\bea
\vec{\partial_b} \tensor{C}(\vec{b} \wedge \vec{a}) = 0 .
\eea

\section{Derivation of the Connection}
\label{connection_derivation}

As discussed in Section~\ref{frames}, a coordinate frame satisfies \eq{torsion_free}:
\bea
	\nabla \wedge \vec{g}^\mu = 0 \qquad \to \qquad \hat{\nabla} \wedge \vec{g}^\mu = - \vec{\partial_a}\wedge (\mv{\Omega}(\vec{a}) \cdot \vec{g}^\mu) .
\label{tf_again}
\eea
In geometric algebra, it is possible to solve \eq{tf_again} for the connection bivectors; we are not aware of a solution in any other language.

To isolate $\mv{\Omega}(\vec{a})$, we first ``differentiate'' out $\vec{g}^\mu$:
\bea
	\vec{g}_\mu \wedge \hat{\nabla} \wedge \vec{g}^\mu = - \vec{g}_\mu \wedge \vec{\partial_a}\wedge (\mv{\Omega}(\vec{a}) \cdot \vec{g}^\mu) = \vec{\partial_a} \wedge \left( \vec{g}_\mu \wedge (\mv{\Omega}(\vec{a}) \cdot \vec{g}^\mu) \right) = - 2 \vec{\partial_a} \wedge \mv{\Omega}(\vec{a}) .
\label{step_1}
\eea
After noting that
\bea
0 = \hat{\nabla} \wedge \left(\vec{g}^\nu \wedge \vec{g}_\nu \right) = \vec{g}_\mu \wedge \hat{\nabla} \wedge \vec{g}^\mu - \vec{g}^\mu \wedge \hat{\nabla} \wedge \vec{g}_\mu ,
\eea
we proceed by taking the inner product of the left and right sides of \eq{step_1} with an arbitrary vector $\vec{b}$:
\begin{subequations}
\bea
	\vec{b} \cdot \left( \vec{g}_\mu \wedge \hat{\nabla} \wedge \vec{g}^\mu \right) &=& \vec{b} \cdot \left( \vec{g}^\mu \wedge \hat{\nabla} \wedge \vec{g}_\mu \right) = b^\mu \hat{\nabla} \wedge \vec{g}_\mu - \vec{g}^\mu \wedge \left( \vec{b} \cdot \hat{\nabla} \vec{g}_\mu \right) + \vec{g}^\mu \wedge \vec{g}^\lambda \vec{b} \cdot \left( \hat{\nabla}_\lambda \vec{g}_\mu \right) \nonumber \\
	&=& b^\mu \hat{\nabla} \wedge \vec{g}_\mu - \vec{g}^\mu \wedge \left( \vec{b} \cdot \hat{\nabla} \vec{g}_\mu \right) + \vec{g}^\mu \wedge \hat{\nabla} b_\mu + \hat{\nabla} \wedge \vec{b} \\
	\vec{b} \cdot \left[ \vec{\partial_a}\wedge\mv{\Omega}(\vec{a}) \right] &=& \mv{\Omega}(\vec{b}) + \vec{\partial_a}\wedge \left(\mv{\Omega}(\vec{a}) \cdot \vec{b} \right) = \mv{\Omega}(\vec{b}) + \nabla \wedge \vec{b} - \hat{\nabla} \wedge \vec{b} .
\eea
\label{step_2}
\end{subequations}
(For simplicity, we have introduced the shorthand $\nabla_\lambda \equiv \vec{g}_\lambda \cdot \nabla$, $b^\mu = \vec{b} \cdot \vec{g}^\mu$ and $b_\mu = \vec{b} \cdot \vec{g}_\mu$.)  We can evaluate the covariant curl of $\vec{b}$ by considering its components $\vec{b} = b_\nu \vec{g}^\nu$:
\bea
	\nabla\wedge \vec{b} = \vec{g}^\lambda \wedge \nabla_\lambda \left( b_\nu \vec{g}^\nu \right) = \vec{g}^\lambda \wedge \left\{ ( \hat{\nabla}_\lambda b_\nu ) \vec{g}^\nu + b_\nu \nabla_\lambda \vec{g}^\nu \right\} = - \vec{g}^\nu \wedge \hat{\nabla} b_\nu
\eea
due to the torsion-free condition and the fact that the components are scalars.

Solving \eq{step_2} for $\mv{\Omega}(\vec{b})$ yields
\bea
	2 \mv{\Omega}(\vec{b}) = \hat{\nabla} \wedge \vec{b} + \vec{g}^\mu \wedge \hat{\nabla} b_\mu - b^\mu \hat{\nabla} \wedge \vec{g}_\mu + \vec{g}^\mu \wedge \left( \vec{b}\cdot \hat{\nabla} \vec{g}_\mu \right) .
\label{step_3}
\eea
This expression is further simplified by unwrapping the curl:
\bea
	\hat{\nabla} \wedge \vec{b} = \vec{g}^\lambda \wedge \hat{\nabla}_\lambda \left( b^\mu \vec{g}_\mu \right) = b^\mu \hat{\nabla} \wedge \vec{g}_\mu - \vec{g}_\mu \wedge \hat{\nabla} b^\mu .
\eea
When plugged into \eq{step_3}, all of this gives us our final expression:
\bea
	\mv{\Omega}(\vec{b}) &=& \frac{1}{2} \left( \vec{g}^\mu \wedge \hat{\nabla} b_\mu - \vec{g}_\mu \wedge \hat{\nabla} b^\mu + \vec{g}^\mu \wedge \left( \vec{b}\cdot \hat{\nabla} \vec{g}_\mu \right) \right) \nonumber \\ 
&=& \frac{1}{2} \left\{ \vec{g}^\mu \wedge \vec{g}^\lambda \vec{b} \cdot \left( \hat{\nabla}_\lambda \vec{g}_\mu \right) - \vec{g}_\mu \wedge \vec{g}^\lambda \vec{b} \cdot \left( \hat{\nabla}_\lambda \vec{g}^\mu \right) + \vec{g}^\mu \wedge \left( \vec{b} \cdot \hat{\nabla} \vec{g}_\mu \right) \right\}.
\label{final_expression}
\eea
The second form of this demonstrates the linearity in $\vec{b}$, while the first is useful in the case $\vec{b} = \vec{g}_\alpha$, where the components are $b^\mu = \delta^\mu_\alpha$ and $b_\mu = g_{\mu \alpha}$; here, \eq{final_expression} reduces to \eq{connection_solution}:
\bea
	\mv{\Omega}_\alpha \equiv \mv{\Omega}(\vec{g}_\alpha ) = \frac{1}{2} \left( \vec{g}^\nu \wedge \hat{\nabla} g_{\alpha \nu} + \vec{g}^\mu \wedge \hat{\nabla}_\alpha \vec{g}_\mu \right) .
\eea
This last equation is derived in Snygg \cite{snygg} using Christoffel symbols and Ricci coefficients, while the more general form is used in Doran \cite{ks1} without referring directly to a coordinate frame.

\section{Null Rotations}
\label{null_rot}

In a physical problem, we may have an obvious null direction specified; to work out the symmetries of the problem, it would be useful to perform an operation that preserves that direction, but changes all the others.  A null Lorentz transformation is the operation we are looking for, and is defined in $\Cl_{1,3}$ by
\bea
\op{\Lambda} (\mv{A}) = \mv{R} \mv{A} \tilde{\mv{R}}
\eea
where
\bea
\mv{R} = \mathrm{e}^{\mv{f}/2 } = 1 + \frac{1}{2} \mv{f},
\eea
and $\mv{f}^2 = 0$ is a bivector.  To preserve the null vector $\vec{l} = 2 \pv{l} \hvec{t}$ (and hence $\pv{l}$),
\bea
\mv{f} = \vec{p} \wedge \vec{l} \qquad \text{where} \quad \vec{p} \cdot \vec{l} = 0
\eea
and $\vec{p}$ is an undetermined spacelike vector.  Due to the orthogonality condition, only two components of this vector need to be determined, which we will write as the complex component of the bivector $\mv{f}$:
\bea
\mv{f} = \sqrt{2} \alpha \mv{Z}_0 = 2 \grade{\pv{p} \pv{l}}_B
\label{fbiv}
\eea
so that $\alpha = p_u - i p_v$ in terms of the bivector basis $\{ \mv{e},\mv{u},\mv{v} \}$.  Under the null rotation, we have a new set of tetrads \cite{exact}:
\bea
\pv{l}' &=& \pv{l} \nonumber \\
\pv{n}' &=& \pv{n} - \alpha \pv{m} - \alpha^\dagger \pv{m}^\dagger + \alpha \alpha^\dagger \pv{l} \nonumber \\
\pv{m}' &=& \pv{m} - \alpha^\dagger \pv{l}.
\label{tetradp}
\eea
Under this same transformation, the bivector basis becomes
\bea
\mv{Z}'_0 &=& - \sqrt{2} \grade{\pv{l}'\pv{m}'}_B = \mv{Z}_0 \nonumber \\
\mv{Z}'_1 &=& \grade{\pv{l}'\pv{n}'}_B - \grade{\pv{m}'\pv{m}'{}^\dagger}_B = \mv{Z}_1 + \sqrt{2} \alpha \mv{Z}_0 \nonumber \\
\mv{Z}'_2 &=& - \sqrt{2} \grade{\pv{n}'\pv{m}'{}^\dagger}_B = \mv{Z}_2 - \sqrt{2} \alpha \mv{Z}_1 - \alpha^2 \mv{Z}_0 .
\label{Zprime}
\eea
In this paper, we will not need the complementary transformation, in which $\pv{n}$ is preserved; however, it is simple to obtain those relations using an analogous discussion \cite{exact}.

Under this transformation, the Weyl matrix elements as defined in \eq{Psi_matrix} change as follows:
\begin{subequations}
\bea
\Psi'_0 &=& -\grade{\mv{Z}'_0 \tensor{C}(\mv{Z}'_0)}_S = \Psi_0 \\
\Psi'_1 &=& -\grade{\mv{Z}'_0 \tensor{C}(\mv{Z}'_1)}_S = \Psi_1 + \sqrt{2} \alpha \Psi_0 \\
\Psi'_2 &=& - \grade{\mv{Z}'_1 \tensor{C}(\mv{Z}'_1)}_S + \grade{\mv{Z}'_0 \tensor{C}(\mv{Z})'_2}_S = \Psi_2 + 3 \sqrt{2} \alpha \Psi_1 + 3 \alpha^2 \Psi_0 \\
\Psi'_3 &=& - \grade{\mv{Z}'_1 \tensor{C}(\mv{Z}'_2)}_S = \Psi_3 - \sqrt{2} \alpha \Psi_2 - 3 \alpha^2 \Psi_1 - \sqrt{2} \alpha^3 \Psi_0 \\
\Psi'_4 &=& -\grade{\mv{Z}'_2 \tensor{C}(\mv{Z}'_2)}_S = \Psi_4 - 2 \sqrt{2} \alpha \Psi_3 + 2 \alpha^2 \Psi_2 + 2 \sqrt{2} \alpha^3 \Psi_1 + \alpha^4 \Psi_0 \label{psi_4_rotation}
\eea
\label{psi_rotations}
\end{subequations}
where the $\mv{Z}'_a$ are given in \eq{Zprime} and $\alpha$ is the complex scalar defined in \eq{fbiv}.  Using transformations like this (and ones that preserve $\pv{n}$), we can adapt our system of tetrads to seek out the Petrov classification for an arbitrary geometry.  As noted in \cite{exact}, we can also find the classification of a known tensor by solving the equation $\Psi'_4 = 0$, which is an algebraic equation in $\alpha$.

\end{appendix}

\bibliography{technique}

\begin{thebibliography}{36}
\expandafter\ifx\csname natexlab\endcsname\relax\def\natexlab#1{#1}\fi
\expandafter\ifx\csname bibnamefont\endcsname\relax
  \def\bibnamefont#1{#1}\fi
\expandafter\ifx\csname bibfnamefont\endcsname\relax
  \def\bibfnamefont#1{#1}\fi
\expandafter\ifx\csname citenamefont\endcsname\relax
  \def\citenamefont#1{#1}\fi
\expandafter\ifx\csname url\endcsname\relax
  \def\url#1{\texttt{#1}}\fi
\expandafter\ifx\csname urlprefix\endcsname\relax\def\urlprefix{URL }\fi
\providecommand{\bibinfo}[2]{#2}
\providecommand{\eprint}[2][]{\url{#2}}

\bibitem[{\citenamefont{Hestenes}(1966)}]{STA}
\bibinfo{author}{\bibfnamefont{D.}~\bibnamefont{Hestenes}},
  \emph{\bibinfo{title}{Space-Time Algebra}} (\bibinfo{publisher}{Gordon and
  Breach}, \bibinfo{address}{New York}, \bibinfo{year}{1966}).

\bibitem[{\citenamefont{Lounesto}(1997)}]{lounesto}
\bibinfo{author}{\bibfnamefont{P.}~\bibnamefont{Lounesto}},
  \emph{\bibinfo{title}{Clifford Algebras and Spinors}}
  (\bibinfo{publisher}{Cambridge University Press},
  \bibinfo{address}{Cambridge}, \bibinfo{year}{1997}).

\bibitem[{\citenamefont{Hestenes and Sobczyk}(1984)}]{CA2GC}
\bibinfo{author}{\bibfnamefont{D.}~\bibnamefont{Hestenes}} \bibnamefont{and}
  \bibinfo{author}{\bibfnamefont{G.}~\bibnamefont{Sobczyk}},
  \emph{\bibinfo{title}{Clifford Algebra to Geometric Calculus: A Unified
  Language for Mathematics and Physics}} (\bibinfo{publisher}{D. Reidel Publ.
  Co.}, \bibinfo{address}{Dordrecht}, \bibinfo{year}{1984}).

\bibitem[{\citenamefont{Sobczyk}(1981{\natexlab{a}})}]{STAC}
\bibinfo{author}{\bibfnamefont{G.}~\bibnamefont{Sobczyk}}, \bibinfo{journal}{J.
  Math. Phys.} \textbf{\bibinfo{volume}{22}}, \bibinfo{pages}{333}
  (\bibinfo{year}{1981}{\natexlab{a}}).

\bibitem[{\citenamefont{Lasenby et~al.}(1997)\citenamefont{Lasenby, Doran, and
  Gull}}]{GTG}
\bibinfo{author}{\bibfnamefont{A.}~\bibnamefont{Lasenby}},
  \bibinfo{author}{\bibfnamefont{C.}~\bibnamefont{Doran}}, \bibnamefont{and}
  \bibinfo{author}{\bibfnamefont{S.}~\bibnamefont{Gull}},
  \bibinfo{journal}{Phil. Trans. R. Soc. London A.}  (\bibinfo{year}{1997}).

\bibitem[{\citenamefont{Wald}(1984)}]{wald}
\bibinfo{author}{\bibfnamefont{R.~M.} \bibnamefont{Wald}},
  \emph{\bibinfo{title}{General Relativity}} (\bibinfo{publisher}{Univ. of
  Chicago Press}, \bibinfo{address}{Chicago}, \bibinfo{year}{1984}).

\bibitem[{\citenamefont{Hestenes}(1998)}]{SCGT}
\bibinfo{author}{\bibfnamefont{D.}~\bibnamefont{Hestenes}}
  (\bibinfo{year}{1998}), \bibinfo{note}{unpublished monograph},
  \urlprefix\url{http://modelingnts.la.asu.edu/pdf/NEW_GRAVITY.pdf}.

\bibitem[{\citenamefont{Choquet-Bruhat
  et~al.}(1982)\citenamefont{Choquet-Bruhat, DeWitt-Morette, and
  Dillard-Bleick}}]{choquet}
\bibinfo{author}{\bibfnamefont{Y.}~\bibnamefont{Choquet-Bruhat}},
  \bibinfo{author}{\bibfnamefont{C.}~\bibnamefont{DeWitt-Morette}},
  \bibnamefont{and}
  \bibinfo{author}{\bibfnamefont{M.}~\bibnamefont{Dillard-Bleick}},
  \emph{\bibinfo{title}{Analysis, Manifolds, and Physics}}
  (\bibinfo{publisher}{North-Holland}, \bibinfo{address}{Amsterdam},
  \bibinfo{year}{1982}), \bibinfo{edition}{revised} ed.

\bibitem[{\citenamefont{Riesz}(1958)}]{riesz}
\bibinfo{author}{\bibfnamefont{M.}~\bibnamefont{Riesz}},
  \emph{\bibinfo{title}{Clifford Numbers and Spinors}}
  (\bibinfo{publisher}{Univ. of Maryland Press}, \bibinfo{address}{College
  Park}, \bibinfo{year}{1958}).

\bibitem[{\citenamefont{Nakahara}(1990)}]{nakahara}
\bibinfo{author}{\bibfnamefont{M.}~\bibnamefont{Nakahara}},
  \emph{\bibinfo{title}{Geometry, Topology and Physics}}
  (\bibinfo{publisher}{Adam Hilger}, \bibinfo{address}{Bristol},
  \bibinfo{year}{1990}).

\bibitem[{\citenamefont{Snygg}(1997)}]{snygg}
\bibinfo{author}{\bibfnamefont{J.}~\bibnamefont{Snygg}},
  \emph{\bibinfo{title}{Clifford Algebra}} (\bibinfo{publisher}{Oxford
  University Press}, \bibinfo{address}{New York}, \bibinfo{year}{1997}).

\bibitem[{\citenamefont{Carroll}(2004)}]{carroll}
\bibinfo{author}{\bibfnamefont{S.~M.} \bibnamefont{Carroll}},
  \emph{\bibinfo{title}{Spacetime and Geometry: An Introduction to General
  Relativity}} (\bibinfo{publisher}{Addison-Wesley}, \bibinfo{address}{San
  Francisco}, \bibinfo{year}{2004}).

\bibitem[{\citenamefont{Martel and Poisson}(2001)}]{martel}
\bibinfo{author}{\bibfnamefont{K.}~\bibnamefont{Martel}} \bibnamefont{and}
  \bibinfo{author}{\bibfnamefont{E.}~\bibnamefont{Poisson}},
  \bibinfo{journal}{Am. J. of Phys.} \textbf{\bibinfo{volume}{69}},
  \bibinfo{pages}{456} (\bibinfo{year}{2001}).

\bibitem[{\citenamefont{Zapolsky}()}]{zapolsky}
\bibinfo{author}{\bibfnamefont{H.}~\bibnamefont{Zapolsky}},
  \bibinfo{note}{private communication}.

\bibitem[{\citenamefont{Heinicke and Hehl}(2001)}]{compgrav}
\bibinfo{author}{\bibfnamefont{C.}~\bibnamefont{Heinicke}} \bibnamefont{and}
  \bibinfo{author}{\bibfnamefont{F.~W.} \bibnamefont{Hehl}}, in
  \emph{\bibinfo{booktitle}{Computer Algebra Handbook}}, edited by
  \bibinfo{editor}{\bibfnamefont{J.}~\bibnamefont{Grabmeier}},
  \bibinfo{editor}{\bibfnamefont{E.}~\bibnamefont{Kaltofen}}, \bibnamefont{and}
  \bibinfo{editor}{\bibfnamefont{V.}~\bibnamefont{Weispfennig}}
  (\bibinfo{publisher}{Springer}, \bibinfo{address}{Berlin},
  \bibinfo{year}{2001}), \eprint{gr-qc/0105094}.

\bibitem[{\citenamefont{Misner et~al.}(1973)\citenamefont{Misner, Thorne, and
  Wheeler}}]{misner}
\bibinfo{author}{\bibfnamefont{C.~W.} \bibnamefont{Misner}},
  \bibinfo{author}{\bibfnamefont{K.~S.} \bibnamefont{Thorne}},
  \bibnamefont{and} \bibinfo{author}{\bibfnamefont{J.~A.}
  \bibnamefont{Wheeler}}, \emph{\bibinfo{title}{Gravitation}}
  (\bibinfo{publisher}{W. H. Freeman and Co.}, \bibinfo{address}{New York},
  \bibinfo{year}{1973}).

\bibitem[{\citenamefont{Hestenes}(1993)}]{GAdiff}
\bibinfo{author}{\bibfnamefont{D.}~\bibnamefont{Hestenes}}, in
  \emph{\bibinfo{booktitle}{Clifford Algebras and Their Applications in
  Mathematical Physics}}, edited by
  \bibinfo{editor}{\bibfnamefont{F.}~\bibnamefont{Brackx}},
  \bibinfo{editor}{\bibfnamefont{R.}~\bibnamefont{Delanghe}}, \bibnamefont{and}
  \bibinfo{editor}{\bibfnamefont{H.}~\bibnamefont{Serras}}
  (\bibinfo{publisher}{Kluwer}, \bibinfo{address}{Dordrecht},
  \bibinfo{year}{1993}), pp. \bibinfo{pages}{269--285}.

\bibitem[{\citenamefont{Kramer et~al.}(1980)\citenamefont{Kramer, Stephani,
  Herlt, and MacCallum}}]{exact}
\bibinfo{author}{\bibfnamefont{D.}~\bibnamefont{Kramer}},
  \bibinfo{author}{\bibfnamefont{H.}~\bibnamefont{Stephani}},
  \bibinfo{author}{\bibfnamefont{E.}~\bibnamefont{Herlt}}, \bibnamefont{and}
  \bibinfo{author}{\bibfnamefont{M.}~\bibnamefont{MacCallum}},
  \emph{\bibinfo{title}{Exact Solutions of Einstein's Field Equations}}
  (\bibinfo{publisher}{Cambridge University Press},
  \bibinfo{address}{Cambridge}, \bibinfo{year}{1980}).

\bibitem[{\citenamefont{Newman and Penrose}(1962)}]{newman-penrose}
\bibinfo{author}{\bibfnamefont{E.}~\bibnamefont{Newman}} \bibnamefont{and}
  \bibinfo{author}{\bibfnamefont{R.}~\bibnamefont{Penrose}},
  \bibinfo{journal}{J. of Math. Phys.} \textbf{\bibinfo{volume}{3}},
  \bibinfo{pages}{566} (\bibinfo{year}{1962}).

\bibitem[{\citenamefont{Janis and Newman}(1965)}]{janis-newman}
\bibinfo{author}{\bibfnamefont{A.~I.} \bibnamefont{Janis}} \bibnamefont{and}
  \bibinfo{author}{\bibfnamefont{E.~T.} \bibnamefont{Newman}},
  \bibinfo{journal}{J. of Math. Phys.} \textbf{\bibinfo{volume}{6}},
  \bibinfo{pages}{902} (\bibinfo{year}{1965}).

\bibitem[{\citenamefont{Penrose and Rindler}(1984)}]{penrose_rindler1}
\bibinfo{author}{\bibfnamefont{R.}~\bibnamefont{Penrose}} \bibnamefont{and}
  \bibinfo{author}{\bibfnamefont{W.}~\bibnamefont{Rindler}},
  \emph{\bibinfo{title}{Spinors and Space-Time, Vol. 1: Two-spinor Calculus and
  Relativistic Fields}} (\bibinfo{publisher}{Cambridge University Press},
  \bibinfo{address}{Cambridge}, \bibinfo{year}{1984}).

\bibitem[{\citenamefont{Baylis}(1999)}]{baylis}
\bibinfo{author}{\bibfnamefont{W.~E.} \bibnamefont{Baylis}},
  \emph{\bibinfo{title}{Electrodynamics: A Modern Geometric Approach}}
  (\bibinfo{publisher}{Birkh{\"a}user}, \bibinfo{address}{Boston},
  \bibinfo{year}{1999}).

\bibitem[{\citenamefont{Sobczyk}(1981{\natexlab{b}})}]{complex}
\bibinfo{author}{\bibfnamefont{G.}~\bibnamefont{Sobczyk}},
  \bibinfo{journal}{Phys. Lett. A} \textbf{\bibinfo{volume}{84}},
  \bibinfo{pages}{45} (\bibinfo{year}{1981}{\natexlab{b}}).

\bibitem[{\citenamefont{Jones and Baylis}(1995)}]{jones}
\bibinfo{author}{\bibfnamefont{G.}~\bibnamefont{Jones}} \bibnamefont{and}
  \bibinfo{author}{\bibfnamefont{W.~E.} \bibnamefont{Baylis}}, in
  \emph{\bibinfo{booktitle}{Clifford Algebras and Spinor Structures}}, edited
  by \bibinfo{editor}{\bibfnamefont{R.}~\bibnamefont{Ab{\l}amowicz}}
  \bibnamefont{and} \bibinfo{editor}{\bibfnamefont{P.}~\bibnamefont{Lounesto}}
  (\bibinfo{publisher}{Kluwer Academic Publishers},
  \bibinfo{address}{Dordrecht}, \bibinfo{year}{1995}), pp.
  \bibinfo{pages}{125--132}.

\bibitem[{\citenamefont{Hestenes}(1967)}]{realspinors}
\bibinfo{author}{\bibfnamefont{D.}~\bibnamefont{Hestenes}},
  \bibinfo{journal}{J. Math. Phys.} \textbf{\bibinfo{volume}{8}},
  \bibinfo{pages}{798} (\bibinfo{year}{1967}).

\bibitem[{\citenamefont{Sobczyk}(1981{\natexlab{c}})}]{classification}
\bibinfo{author}{\bibfnamefont{G.}~\bibnamefont{Sobczyk}},
  \bibinfo{journal}{Phys. Lett. A} \textbf{\bibinfo{volume}{84}},
  \bibinfo{pages}{49} (\bibinfo{year}{1981}{\natexlab{c}}).

\bibitem[{\citenamefont{Hestenes}(1985)}]{classical}
\bibinfo{author}{\bibfnamefont{D.}~\bibnamefont{Hestenes}},
  \emph{\bibinfo{title}{New Foundations for Classical Mechanics}}
  (\bibinfo{publisher}{D. Reidel Publ. Co.}, \bibinfo{address}{Dordrecht},
  \bibinfo{year}{1985}).

\bibitem[{\citenamefont{Szekeres}(1965)}]{compass}
\bibinfo{author}{\bibfnamefont{P.}~\bibnamefont{Szekeres}},
  \bibinfo{journal}{J. of Math. Phys.} \textbf{\bibinfo{volume}{6}},
  \bibinfo{pages}{1387} (\bibinfo{year}{1965}).

\bibitem[{\citenamefont{Doran et~al.}()\citenamefont{Doran, Lasenby, and
  Gull}}]{ks2}
\bibinfo{author}{\bibfnamefont{C.}~\bibnamefont{Doran}},
  \bibinfo{author}{\bibfnamefont{A.}~\bibnamefont{Lasenby}}, \bibnamefont{and}
  \bibinfo{author}{\bibfnamefont{S.}~\bibnamefont{Gull}},
  \bibinfo{note}{unpublished},
  \urlprefix\url{http://www.mrao.cam.ac.uk/~clifford/publications/abstracts/kerr_schild2.html}.

\bibitem[{\citenamefont{Schiffer et~al.}(1973)\citenamefont{Schiffer, Adler,
  Mark, and Sheffield}}]{schiffer}
\bibinfo{author}{\bibfnamefont{M.~M.} \bibnamefont{Schiffer}},
  \bibinfo{author}{\bibfnamefont{R.~J.} \bibnamefont{Adler}},
  \bibinfo{author}{\bibfnamefont{J.}~\bibnamefont{Mark}}, \bibnamefont{and}
  \bibinfo{author}{\bibfnamefont{C.}~\bibnamefont{Sheffield}},
  \bibinfo{journal}{J. Math. Phys.} \textbf{\bibinfo{volume}{14}},
  \bibinfo{pages}{52} (\bibinfo{year}{1973}).

\bibitem[{\citenamefont{Doran}()}]{ks1}
\bibinfo{author}{\bibfnamefont{C.}~\bibnamefont{Doran}},
  \bibinfo{note}{unpublished},
  \urlprefix\url{http://www.mrao.cam.ac.uk/~clifford/publications/abstracts/kerr_schild1.html}.

\bibitem[{\citenamefont{Chandrasekhar}(1992)}]{chandra}
\bibinfo{author}{\bibfnamefont{S.}~\bibnamefont{Chandrasekhar}},
  \emph{\bibinfo{title}{The Mathematical Theory of Black Holes}}
  (\bibinfo{publisher}{Oxford University Press}, \bibinfo{address}{Oxford},
  \bibinfo{year}{1992}).

\bibitem[{\citenamefont{Sobczyk}(1986)}]{killing}
\bibinfo{author}{\bibfnamefont{G.~E.} \bibnamefont{Sobczyk}}, in
  \emph{\bibinfo{booktitle}{Clifford Algebras and Their Applications in
  Mathematical Physics}}, edited by \bibinfo{editor}{\bibfnamefont{J.~S.~R.}
  \bibnamefont{Chisholm}} \bibnamefont{and}
  \bibinfo{editor}{\bibfnamefont{A.~K.} \bibnamefont{Common}}
  (\bibinfo{publisher}{D. Reidel}, \bibinfo{address}{Dordrecht},
  \bibinfo{year}{1986}).

\bibitem[{\citenamefont{Lasenby et~al.}(1993)\citenamefont{Lasenby, Doran, and
  Gull}}]{2spinors}
\bibinfo{author}{\bibfnamefont{A.}~\bibnamefont{Lasenby}},
  \bibinfo{author}{\bibfnamefont{C.}~\bibnamefont{Doran}}, \bibnamefont{and}
  \bibinfo{author}{\bibfnamefont{S.}~\bibnamefont{Gull}}, in
  \emph{\bibinfo{booktitle}{Spinors, Twistors, Clifford Algebras and Quantum
  Deformations}}, edited by
  \bibinfo{editor}{\bibfnamefont{A.}~\bibnamefont{Borowiec}} \bibnamefont{and}
  \bibinfo{editor}{\bibfnamefont{B.}~\bibnamefont{Jancewicz}}
  (\bibinfo{publisher}{Kluwer Academic}, \bibinfo{address}{Dordrecht},
  \bibinfo{year}{1993}), p. \bibinfo{pages}{233}.

\bibitem[{\citenamefont{Gull et~al.}(1993)\citenamefont{Gull, Lasenby, and
  Doran}}]{imag_numbs}
\bibinfo{author}{\bibfnamefont{S.}~\bibnamefont{Gull}},
  \bibinfo{author}{\bibfnamefont{A.}~\bibnamefont{Lasenby}}, \bibnamefont{and}
  \bibinfo{author}{\bibfnamefont{C.}~\bibnamefont{Doran}},
  \bibinfo{journal}{Found. Phys.} \textbf{\bibinfo{volume}{23}},
  \bibinfo{pages}{1175} (\bibinfo{year}{1993}).

\bibitem[{\citenamefont{Hestenes}(2003)}]{intro_GA}
\bibinfo{author}{\bibfnamefont{D.}~\bibnamefont{Hestenes}},
  \bibinfo{journal}{Am. J. Phys.} \textbf{\bibinfo{volume}{71}},
  \bibinfo{pages}{1} (\bibinfo{year}{2003}).

\end{thebibliography}

\end{document}